\documentclass[aps,prd,twocolumn,groupedaddress,superscriptaddress]{revtex4-2}

\usepackage{lipsum}
\usepackage{graphicx}  
\usepackage{dcolumn}   
\usepackage{bm}        
\usepackage{amssymb}   
\usepackage{amsmath}
\usepackage{xcolor}
\usepackage{amsmath}
\usepackage{amsfonts}
\usepackage{hyperref}

\usepackage{float}

\usepackage{hyperref}
\hypersetup{colorlinks=true, linkcolor=blue, citecolor=blue, urlcolor=blue}

\usepackage[normalem]{ulem}

\begin{document}
\preprint{APS/123-QED}

\title{Bayesian Inference of the Critical Endpoint in 2+1-Flavor System \\from Holographic QCD}

\author{Liqiang Zhu}
\email{zhuliqiang@mails.ccnu.edu.cn}
\affiliation{Key Laboratory of Quark and Lepton Physics (MOE) and Institute of Particle Physics, Central China Normal University, Wuhan 430079, China}

\author{Xun Chen}
\email{chenxun@usc.edu.cn}
\affiliation{Key Laboratory of Quark and Lepton Physics (MOE) and Institute of Particle Physics, Central China Normal University, Wuhan 430079, China}
\affiliation{School of Nuclear Science and Technology, University of South China, Hengyang 421001, China}
\affiliation{INFN -- Istituto Nazionale di Fisica Nucleare -- Sezione di Bari, Via Orabona 4, 70125 Bari, Italy}

\author{Kai Zhou}
\email{zhoukai@cuhk.edu.cn}
\affiliation{School of Science and Engineering, The Chinese University of Hong Kong, Shenzhen (CUHK-Shenzhen), Guangdong, 518172, China}
\affiliation{Frankfurt Institute for Advanced Studies, Ruth Moufang Strasse 1, D-60438, Frankfurt am Main, Germany}

\author{Hanzhong Zhang}
\email{zhanghz@mail.ccnu.edu.cn}
\affiliation{Key Laboratory of Quark and Lepton Physics (MOE) and Institute of Particle Physics, Central China Normal University, Wuhan 430079, China}

\author{Mei Huang}
\email{huangmei@ucas.ac.cn}
\affiliation{School of Nuclear Science and Technology, University of Chinese Academy of Sciences, Beijing 100049, China}

\begin{abstract}
We present a Bayesian holographic model constructed by integrating the equation of state and baryon number susceptibility at zero chemical potential from lattice QCD. The model incorporates error estimates derived from lattice data. With this model, we systematically investigate the thermodynamic properties of the 2+1-flavor  
QCD system. Using Bayesian Inference, we perform precise calibration of the model parameters and determined the critical endpoint (CEP) position under the maximum a posterior (MAP) estimation to be $(T^{c},\mu_{B}^{c})=(0.0859\;\mathrm{GeV},0.742\;\mathrm{GeV})$. Additionally, we predict the CEP positions within 68\% and 95\% confidence levels, yielding $(T^{c},\; \mu_{B}^{c})_{68\%}$=$(0.0820\text{--}0.0889,
0.71\text{--}0.77)\;\mathrm{GeV}$ and $(T^{c},\; \mu_{B}^{c})_{95\%}$=$(0.0816\text{--}0.0898,\;
0.71\text{--}0.79)\;\mathrm{GeV}$, respectively. Moreover, to validate the reliability and predictive power of our approach, we conduct a comprehensive comparison between our predictions and potential CEP locations proposed by other theoretical models.  This work not only establishes a novel Bayesian framework for holographic modeling but also provides valuable insights and theoretical support for exploring phase transitions in strongly-interacting matter under extreme conditions.
\end{abstract}
\maketitle

\section{Introduction}
\label{secintro}

In the context of the Standard Model, the strong interaction among quarks and gluons is described by Quantum Chromodynamics (QCD). A fundamental prediction of QCD based upon its asymptotic properties is that, at sufficiently high temperatures or densities, strongly interacting matter transitions from confined hadronic state into a novel deconfined state called quark-gluon plasma (QGP)\cite{PHENIX:2004vcz, STAR:2010vob, PHENIX:2001hpc, STAR:2003fka, Xie:2024xbn, Li:2024uzk}.
In this new state, quarks and gluons are no longer confined within hadrons but instead can interact and move freely. This new state represent the behavior of QCD matter under extreme conditions. The exploration of this transition, and more generally of the QCD phase diagram, has been a longstanding pursuit in high energy physics and remains a central area of inquiry 
\cite{Fu:2022gou,Guenther:2020jwe,Zhou:2023pti,Du:2024wjm,Luo:2020pef, Ma:2023zfj}.

Theoretically, the primary method for investigating the QCD phase diagram relies on first-principle lattice calculations. While the equation of state (EoS) can be reliably obtained at finite temperatures at zero and small chemical potentials, the inherent sign problem in lattice simulations presents obstacles when applying the simulation to situations involving finite baryon density. To gain deeper insights into the QCD phase diagram, a variety of alternative theoretical frameworks have also been employed  including the Functional Renormalization Group (FRG) \cite{Fu:2022gou,Zhang:2017icm}, the Dyson-Schwinger equations (DSEs) \cite{Gao:2016qkh,Qin:2010nq,Shi:2014zpa}, and effective QCD models including the Random Matrix Model (RMM) \cite{Halasz:1998qr}, the Nambu–Jona-Lasinio (NJL) models \cite{Berges:1998rc,Stephanov:2004wx,Li:2018ygx,Sun:2024anu,Bao:2024glw}. These approaches provide a rich array of theoretical tools and perspectives for understanding the behavior of strongly interacting matter.

In recent decades, the AdS/CFT correspondence \cite{Maldacena:1997re} has become an increasingly important non-perturbative framework for studying both hadronic physics and hot/dense QCD matter. This theoretical framework not only offers profound insights into hadrons \cite{Erdmenger:2007cm,Brodsky:2014yha,Casalderrey-Solana:2011dxg,Adams:2012th,Li:2016gtz,BitaghsirFadafan:2018uzs,Abt:2019tas,Nakas:2020hyo,Li:2013oda}, but also opens new perspectives for investigating QCD matter under extreme conditions \cite{Gubser:2008ny,DeWolfe:2010he,Jarvinen:2022doa,Yang:2015aia,Yang:2014bqa,Dudal:2017max,Dudal:2018ztm,Fang:2015ytf,Liu:2023pbt,Li:2022erd,Critelli:2017oub,Grefa:2021qvt,Arefeva:2021jpa,Arefeva:2020vae,Chen:2018vty,Chen:2017lsf,Chen:2020ath,Zhou:2020ssi,Chen:2019rez,Cao:2022csq,Arcos:2022icf,Cao:2024jgt,Bohra:2020qom,Arefeva:2023jjh,Bohra:2019ebj}. Among the numerous studies conducted under this framework, 
the Einstein-Maxwell-dilaton (EMD) model \cite{Gubser:2008ny,DeWolfe:2010he,Li:2013oda,Yang:2015aia,Yang:2014bqa,Dudal:2017max,Dudal:2018ztm,Fang:2015ytf,Liu:2023pbt,Li:2022erd,Critelli:2017oub,Grefa:2021qvt,Arefeva:2021jpa,Arefeva:2020vae,Chen:2020ath,Chen:2019rez,Cao:2024jgt} has proven particularly effective at capturing key features of QGP. Concurrently, with the advancement of machine learning techniques, these methods have been applied to holography to construct models that utilize specific experimental and lattice QCD data to determine the bulk metric \cite{Hashimoto:2018ftp,Hashimoto:2018bnb,Akutagawa:2020yeo,Hashimoto:2022eij,Yan:2020wcd,Song:2020agw,Chang:2024ksq,Ahn:2024gjf,Gu:2024lrz,Li:2022zjc,Ahn:2024lkh,Chen:2024ckb,Chen:2024jet}. Unlike conventional holographic models, this data-driven approach offers a powerful avenue for advancing our understanding of the properties of QGP.

Bayesian inference has found extensive applications in heavy-ion physics, addressing several critical analysis such as determining jet transport coefficients, modeling energy loss distributions, estimating bulk viscosity and QCD EoS, etc.,\cite{Xie:2022fak,Xing:2023ciw,Wu:2023azi,Zhu:2025edv,JETSCAPE:2023ikg,He:2018gks,JETSCAPE:2021rfx,Parkkila:2021tqq,Bernhard:2019bmu, Cheng:2023ucp}. Additionally, it employs global fitting techniques to estimate parameters from a diverse range of observables \cite{Pratt:2015zsa,Bernhard:2016tnd,Bernhard:2019bmu,Du:2019civ}. Recently, in Ref. \cite{Hippert:2023bel}, Bayesian inference was applied to the EMD model, where specific functional forms for the dilaton potential $V(\phi)$ and the coupling between the dilaton and gauge fields $f(\phi)$ were assumed. In this study, we adopt the potential reconstruction method, where one can input either the dilaton or a metric to determine the dilaton potential, and then incorporate Bayesian inference into the EMD model. A significant advantage of the potential reconstruction method in our model is that it enables analytical solutions, thus offering robust theoretical support for subsequent investigation over different interesting physics.

Compared to the machine learning-based EMD model \cite{Chen:2024ckb,Chen:2024mmd}, a key feature of this work is the systematic incorporation of error bars derived from lattice QCD data, allowing us to provide confidence intervals for the derived physics rather than relying solely on best-fit estimates. This approach enhances the reliability and robustness of our results. The structure of the remainder of this paper is organized as follows: Section. \ref{sec2} provides a detailed review of the EMD model, outlining its fundamental principles and application background; Section \ref{sec3} delves into our specific procedures of parameter estimation using Bayesian inference, illustrating how the lattice data including their error estimation can be utilized to effectively calibrate our model with more robustness; In Section \ref{sec4}, we thoroughly present various thermodynamic quantities derived from our Bayesian calibrated holographic model, including energy density and pressure; Section \ref{sec5} focuses on the critical endpoint (CEP) positions predicted by our model, comparing these findings with those of other theoretical frameworks; Lastly, Section \ref{sec6} concludes the paper by summarizing the main results and suggesting directions for future research.

\section{The review of holographic EMD model}
\label{sec2}

In this section, we provide a comprehensive review of the five-dimensional EMD system from the perspective of potential reconstruction \cite{Li:2011hp,Cai:2012xh,He:2013qq,Yang:2014bqa,Yang:2015aia,Dudal:2017max,Dudal:2018ztm,Chen:2018vty,Chen:2020ath,Zhou:2020ssi,Chen:2019rez,Lin:2024mct}. The action of the system incorporates a gravitational field $g_{\mu \nu}^s$, a Maxwell field $A_\mu$,  and a neutral dilaton scalar field. Within the string frame, its dynamical behavior is characterized by the following equation
\begin{equation}
\begin{split}
S_b&=\frac{1}{16 \pi G_5} \int d^5 x \sqrt{-g^s} e^{-2 \phi_s}\\
&\quad\times\left[R_s-\frac{f_s\left(\phi_s\right)}{4} F^2+4 \partial_\mu \phi_s \partial^\mu \phi_s-V_s\left(\phi_s\right)\right],
\label{Eq:actionsb}
\end{split}
\end{equation}
Here, $f(\phi)$ denotes the gauge kinetic function that interacts with the Maxwell field $A_\mu$, while $V\left(\phi\right)$ signifies the potential of the dilaton field, and $G_5$ refers to the Newton constant in a five-dimensional spacetime. The specific forms of $f\left(\phi\right)$ and $V\left(\phi\right)$ can be systematically identified by solving the equations of motion (EoMs). To delve deeper into the thermodynamic characteristics of quantum chromodynamics (QCD), we perform a transformation of the action from the string frame to the Einstein frame based on the following relations,
\begin{equation}
\begin{split}
\phi_s=\sqrt{\frac{3}{8}} \phi, \quad &g_{\mu \nu}^s=g_{\mu \nu} e^{\sqrt{\frac{2}{3}} \phi}, \quad f_s\left(\phi_s\right)=f(\phi) e^{\sqrt{\frac{2}{3}} \phi},\\
&V_s\left(\phi_s\right)=e^{-\sqrt{\frac{2}{3}} \phi} V(\phi).
\end{split}
\end{equation}
After transforming to the Einstein frame, the action takes the following form
\begin{equation}
\begin{split}
S_b & =\frac{1}{16 \pi G_5} \int d^5 x \sqrt{-g}\\
&\quad\times\left[R-\frac{f(\phi)}{4} F^2-\frac{1}{2} \partial_\mu \phi \partial^\mu \phi-V(\phi)\right].
\end{split}
\end{equation}
The EoMs obtained from the action are listed as follows
\begin{equation}
\begin{gathered}
R_{M N}-\frac{1}{2} g_{M N} R-T_{M N}=0, \\
\nabla_M\left[f(\phi) F^{M N}\right]=0, \\
\partial_M\left[\sqrt{-g} \partial^M \phi\right]-\sqrt{-g}\left(\frac{\partial V}{\partial \phi}+\frac{F^2}{4} \frac{\partial f}{\partial \phi}\right)=0,
\end{gathered}
\end{equation}
where
\begin{equation}
\begin{aligned}
T_{M N}&=\frac{1}{2}\left(\partial_M \phi \partial_N \phi-\frac{1}{2} g_{M N}(\partial \phi)^2-g_{M N} V(\phi)\right)\\
&\quad+\frac{f(\phi)}{2}\left(F_{M P} F_N^P-\frac{1}{4} g_{M N} F^2\right) .
\end{aligned}
\end{equation}
Here, we propose the following assumed form of the metric
\begin{equation}
d s^2=\frac{L^2 e^{2 A(z)}}{z^2}\left[-g(z) d t^2+\frac{d z^2}{g(z)}+d \vec{x}^2\right],
\end{equation}
In this framework, $z$ represents the holographic coordinate in five dimensions, with the radius $L$ of the $\rm AdS_5$ space normalized to one. Utilizing the proposed metric ansatz, we can derive the EoMs and the corresponding constraints for the background fields, as detailed below,
\begin{equation}\label{7}
\begin{footnotesize}
\begin{gathered}
\phi^{\prime \prime}+\phi^{\prime}\left(-\frac{3}{z}+\frac{g^{\prime}}{g}+3 A^{\prime}\right)-\frac{L^2 e^{2 A}}{z^2 g} \frac{\partial V}{\partial \phi}+\frac{z^2 e^{-2 A} A_t^{\prime 2}}{2 L^2 g} \frac{\partial f}{\partial \phi}=0,
\end{gathered}
\end{footnotesize}
\end{equation}
\begin{equation}\label{8}
A_t^{\prime \prime}+A_t^{\prime}\left(-\frac{1}{z}+\frac{f^{\prime}}{f}+A^{\prime}\right)=0,
\end{equation}
\begin{equation}\label{9}
g^{\prime \prime}+g^{\prime}\left(-\frac{3}{z}+3 A^{\prime}\right)-\frac{e^{-2 A} A_t^{\prime 2} z^2 f}{L^2}=0,
\end{equation}
\begin{equation}\label{10}
\begin{split}
&A^{\prime \prime}+\frac{g^{\prime \prime}}{6 g}+A^{\prime}\left(-\frac{6}{z}+\frac{3 g^{\prime}}{2 g}\right)\\
&-\frac{1}{z}\left(-\frac{4}{z}+\frac{3 g^{\prime}}{2 g}\right)+3 A^{\prime 2} +\frac{L^2 e^{2 A} V}{3 z^2 g}=0,
\end{split}
\end{equation}
\begin{equation}\label{11}
A^{\prime \prime}-A^{\prime}\left(-\frac{2}{z}+A^{\prime}\right)+\frac{\phi^{\prime 2}}{6}=0,
\end{equation}
Among these five equations, only four are linearly independent. 
The horizon ($z=z_{h}$) is the boundary condition for the equations of motion (Eq. \ref{7}-\ref{11}),
and horizon is also the boundary of the black hole, which physically corresponds to the introduction of temperature, and is represented as follows
\begin{equation}
A_t\left(z_h\right)=g\left(z_h\right)=0.
\end{equation}
As we approach the infrared (IR) boundary at $z \rightarrow z_{h}$, we impose that the metric in the string frame approaches $\rm AdS_5$. The ultraviolet (UV) boundary ($z=0$) conditions are as follows
\begin{equation}
\begin{small}
A(0)=-\sqrt{\frac{1}{6}} \phi(0) =0,  g(0)=1,  A_t(0)=\mu+\rho^{\prime} z^2+\cdots.
\end{small}
\end{equation}
The parameter $\mu$ represents the baryon chemical potential, while $\rho^{\prime}$ corresponds to a quantity proportional to the baryon number density. The connection between $\mu$ and the quark chemical potential is expressed as $\mu = 3\mu_q$. The procedure for computing the baryon number density is detailed in \cite{Critelli:2017oub,Zhang:2022uin}:
\begin{equation}
\begin{aligned}
\rho & =\left|\lim _{z \rightarrow 0} \frac{\partial \mathcal{L}}{\partial\left(\partial_z A_t\right)}\right| \\
& =-\frac{1}{16\pi G_5} \lim _{z \rightarrow 0}\left[\frac{\mathrm{e}^{A(z)}}{z} f(\phi) \frac{\mathrm{d}}{\mathrm{d} z} A_t(z)\right].
\end{aligned}
\end{equation}
In the Einstein frame, the Lagrangian density $\mathcal{L}$ enables the explicit determination of the EoMs using analytical methods
\begin{align}
\begin{split}
\phi^{\prime}(z) & =\sqrt{-6\left(A^{\prime \prime}-A^{\prime 2}+\frac{2}{z} A^{\prime}\right)}, \\
A_t(z) & =\sqrt{\frac{-1}{\int_0^{z_h} y^3 e^{-3 A} d y \int_{y_g}^y \frac{x}{e^A f} d x}} \int_{z_h}^z \frac{y}{e^A f} d y, \\
g(z) & =1-\frac{\int_0^z y^3 e^{-3 A} d y \int_{y_g}^y \frac{x}{e^A f} d x}{\int_0^{z_h} y^3 e^{-3 A} d y \int_{y_g}^y \frac{x}{e^A f} d x}, \\
V(z) & =-3 z^2 g e^{-2 A}\bigg[A^{\prime \prime}+3 A^{\prime 2}+\left(\frac{3 g^{\prime}}{2 g}-\frac{6}{z}\right) A^{\prime}\\
& \quad -\frac{1}{z}\left(\frac{3 g^{\prime}}{2 g}-\frac{4}{z}\right)+\frac{g^{\prime \prime}}{6 g}\bigg].
\end{split}
\end{align}
The parameter $y_g$, as the only undetermined constant, may be linked to the chemical potential $\mu$ in a specific manner. This is accomplished by expanding the field $A_t(z)$ in the vicinity of the boundary at $z=0$, resulting in
\begin{equation}
\begin{split}
A_t(0)&=\sqrt{\frac{-1}{\int_0^{z_h} y^3 e^{-3 A} d y \int_{y_g}^y \frac{x}{e^A f} d x}}\\
&\quad\times\bigg(-\int_0^{z_h} \frac{y}{e^A f} d y+\frac{1}{e^{A(0)} f(0)} z^2+\cdots\bigg).
\end{split}
\end{equation}
Following the principles of the AdS/CFT correspondence, the chemical potential of the system can be characterized as:
\begin{equation}
\mu=-\sqrt{\frac{-1}{\int_0^{z_h} y^3 e^{-3 A} d y \int_{y_g}^y \frac{x}{e^A f} d x}} \int_0^{z_h} \frac{y}{e^A f} d y.
\end{equation}
The function form of the gauge kinetic term $f(z)$ is specified as: \cite{Chen:2024ckb,Chen:2024mmd,Chen:2024epd}
\begin{equation}\label{fff}
f(z)=e^{c z^2-A(z)+k}.
\end{equation}
The function $f(z)$ encapsulates the relationship between the model and the chemical potential, guided by the baryon number susceptibility. Additionally, this particular choice of $f(z)$ is made to simplify the derivation of the analytical solution. The connection between the integration constant $y_g$ and the chemical potential $\mu$ is given by:
\begin{equation}
e^{-c y_g^2}=\frac{\int_0^{z_h} y^3 e^{-3 A-c y^2} d y}{\int_0^{z_h} y^3 e^{-3 A} d y}-\frac{\left(1-e^{-c z_h^2}\right)^2}{2 c \mu^2 e^k \int_0^{z_h} y^3 e^{-3 A} d y}.
\end{equation}
From this, we are able to derive
\begin{equation}
\begin{aligned}
g(z)&=1-\frac{1}{\int_0^{z h} d x x^3 e^{-3 A(x)}}\\
&\quad\times\Bigg[\int_0^z d x x^3 e^{-3 A(x)}+\frac{2 c \mu^2 e^k}{\left(1-e^{-c z_h^2}\right)^2} \operatorname{det} \mathcal{G}\Bigg],\\
\phi^{\prime}(z) & =\sqrt{6\left(A^{\prime 2}-A^{\prime \prime}-2 A^{\prime} / z\right)}, \\
A_t(z) & =\mu \frac{e^{-c z^2}-e^{-c z_h^2}}{1-e^{-c z_h^2}}, \\
V(z) & =-\frac{3 z^2 g e^{-2 A}}{L^2}\bigg[A^{\prime \prime}+A^{\prime}\left(3 A^{\prime}-\frac{6}{z}+\frac{3 g^{\prime}}{2 g}\right)\\
&\quad-\frac{1}{z}\left(-\frac{4}{z}+\frac{3 g^{\prime}}{2 g}\right)+\frac{g^{\prime \prime}}{6 g}\bigg],
\end{aligned}
\end{equation}
where
\begin{equation}
\operatorname{det} \mathcal{G}=\left|\begin{array}{ll}
\int_0^{z_h} d y y^3 e^{-3 A(y)} & \int_0^{z_h} d y y^3 e^{-3 A(y)-c y^2} \\
\int_{z_h}^z d y y^3 e^{-3 A(y)} & \int_{z_h}^z d y y^3 e^{-3 A(y)-c y^2}
\end{array}\right|.
\end{equation}
The Hawking temperature ~\cite{Natsuume:2014sfa} and entropy corresponding to this black hole solution can be represented as follows
\begin{equation}
\begin{small}
\begin{aligned}
T & =\frac{\left|g^{\prime}(z)\right|}{4\pi}\\ 
& =\frac{z_h^3 e^{-3 A\left(z_h\right)}}{4 \pi \int_0^{z_h} d y y^3 e^{-3 A(y)}}\Bigg[1+ \\
&\frac{2 c \mu^2 e^k\left(e^{-c z_h^2} \int_0^{z_h} d y y^3 e^{-3 A(y)}-\int_0^{z_h} d y y^3 e^{-3 A(y)} e^{-c y^2}\right)}{(1-e^{-c z_h^2})^2} \Bigg],
\end{aligned}
\end{small}
\end{equation}
\begin{equation}\label{SSS}
S=\frac{e^{3 A\left(z_h\right)}}{4 G_5 z_h^3}.
\end{equation}
In order to solve the system analytically, we make the following ansatz:
\begin{equation}\label{AAA}
A(z)= d*\ln(a z^2 + 1) + d*\ln(b z^4 + 1).
\end{equation}
The function $A(z)$ is designed to reproduce the proper entropy behavior and impose constraints on the temperature-dependent model inspired by Refs. \cite{Dudal:2017max,Chen:2020ath,Chen:2024ckb,Chen:2024mmd}. Within the string frame, $A_s(z)$ is expressed as:
\begin{equation}
A_s(z)=A(z)+\sqrt{\frac{1}{6}} \phi(z).
\end{equation}
The model involves five undetermined parameters, which can be fixed using Bayesian Inference based on the equation of state (EoS) data provided by lattice calculations. Additionally, a parameter $c$ can be concurrently identified through the analysis of baryon number susceptibility. Once the entropy is obtained, the free energy can then be derived as:
\begin{equation}
\begin{aligned}
F&=-\int s d T-\int \rho d \mu\\
&=-\int s\left(\frac{\partial T}{\partial z_h} d z_h+\frac{\partial T}{\partial \mu} d \mu\right)-\int\rho d \mu\\
&= \int_{z_h}^\infty s\frac{\partial T}{\partial z_h} d z_h-\int_0^\mu (\frac{\partial T}{\partial \mu} d \mu\ +\rho) d \mu.
\end{aligned}
\end{equation}
The free energy has been normalized such that it approaches zero as $z_h \rightarrow \infty$. Consequently, The energy density associated with the system can be formulated as:
\begin{equation}
\epsilon=-p+s T+\mu \rho.
\end{equation}
In conditions with a non-zero chemical potential, the squared sound speed can be determined using the approaches outlined in \cite{Li:2020hau,Yang:2017oer,Gursoy:2017wzz}
\begin{equation}
C_s^2=\frac{s}{T\left(\frac{\partial s}{\partial T}\right)_\mu+\mu\left(\frac{\partial \rho}{\partial T}\right)_\mu}.
\end{equation}
Specific heat capacity refers to the amount of heat absorbed or released by a unit mass of a substance during temperature changes, which can be defined as:
\begin{equation}
C_V=T \frac{\partial s}{\partial T}.
\end{equation}
In this work, we define the scaled second-order baryon number susceptibility as:
\begin{equation}
\chi_{2}^{B}=\frac{1}{T^2} \frac{\partial \rho}{\partial \mu}.
\end{equation}

\section{Bayesian inference of model parameters}\label{sec3}

\begin{figure*}
     \centering
     \includegraphics[width=0.7\textwidth]{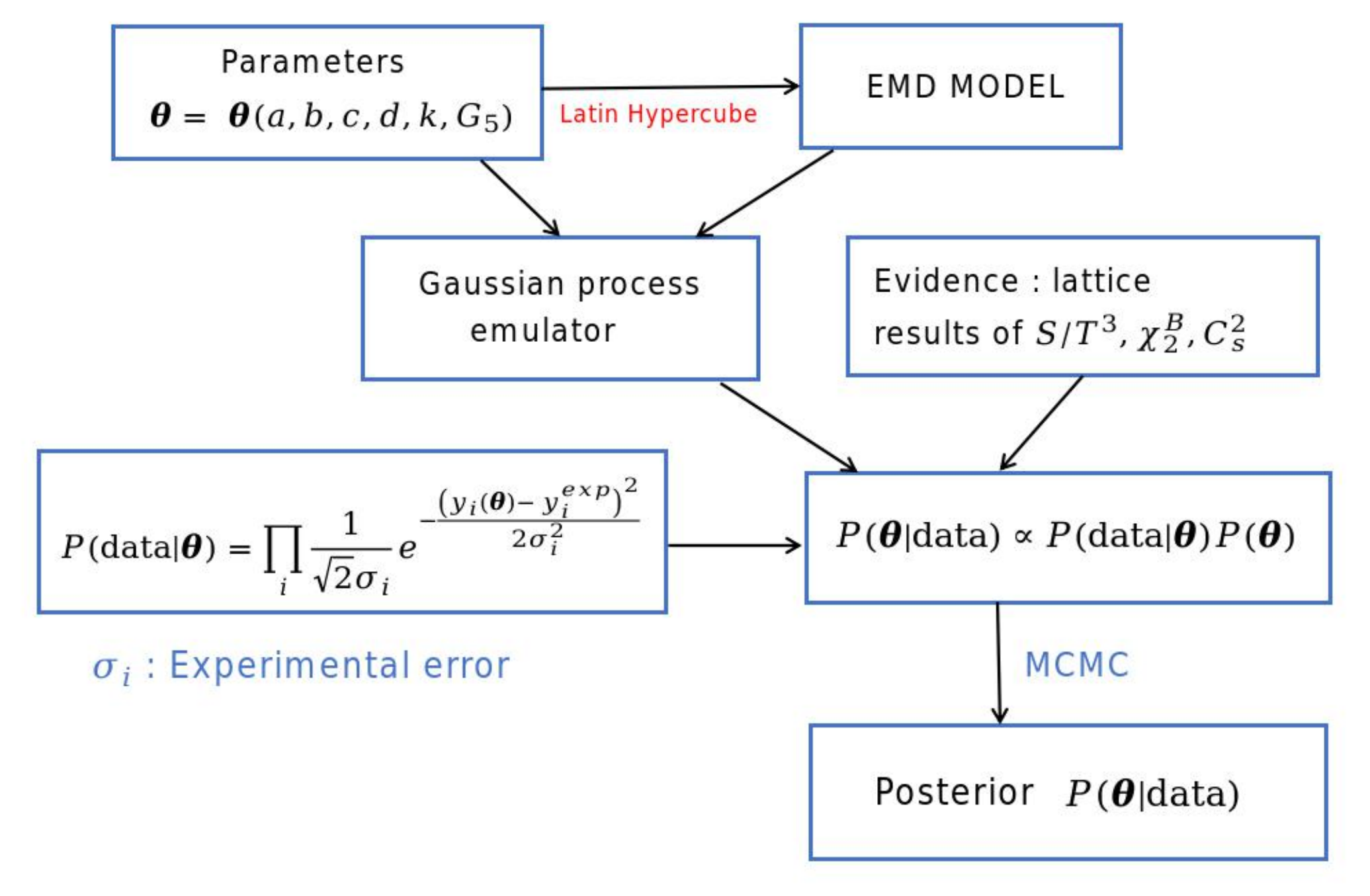}
     \caption{Overview of the Bayesian parameter inference process.}
     \label{fig1}
\end{figure*}

Bayesian inference has proven highly successful in constraining model parameters within relativistic heavy-ion collisions  \cite{Pratt:2015zsa,Novak:2013bqa,Sangaline:2015isa}. Notable applications include determinaitons of the shear viscosity over entropy density ratio ($\eta/s$) of the QGP medium \cite{Bernhard:2015hxa,Bernhard:2016tnd}, the jet transport coeﬃcient $\hat{q}$ \cite{Xie:2022ght,Xie:2022fak,JETSCAPE:2021ehl}, the QCD Equation of State \cite{OmanaKuttan:2022aml}, nucleon distributions in the nucleus \cite{Cheng:2023ucp}, and the analysis of the jet energy loss \cite{He:2018gks,Wu:2023azi, Xing:2023ciw}. In this work, we leverage Bayesian methods to systematically investigate how Lattice QCD results \cite{HotQCD:2014kol,Bazavov:2017dus} constrain the parameter space of the EMD model. Specifically, we infer the posterior distributions of the EMD model parameters $\boldsymbol{\theta}(a,b,c,d,k,G_{5})$ to study their influence on predictions for the thermodynamic behaviors of the entropy $S_{BH}$, baryon susceptibility $\chi_{2}^{B}$ and speed of sound squared $C_{s}^{2}$.  
This section outlines the Bayesian inference framework applied to the EMD model, with a schematic of the workflow provided in Fig. \ref{fig1}. 
Subsection A details the model inputs of the EMD model, including parameter design, selection criteria, as well as their prior distribution. Subsection B describes the model outputs generated from these inputs and the post-processing steps applied to enable comparison with lattice provided observables. 
Subsection C elaborates on the technical implementation  
of the Gaussian process emulator, which bridges the parameter spaces to observables. Finally, subsection D presents specific process of parameter inference, covering the construction of prior distributions, likelihood functions, and posterior distributions, as well as the Markov Chain Monte Carlo (MCMC) \cite{Foreman-Mackey:2012any, Goodman:2010dyf} sampling to explore the posterior distribution.

\subsection{Parameter design}

This subsection outlines the design of the EMD model parameters 
$\boldsymbol{\theta}=(a,b,c,d,k,G_{5})$. The process comprises two essential steps: initially, defining the number of design points and their respective value ranges for each parameter; subsequently, strategically distributing these design points within the parameter space.

First, the number of design points, which corresponds to how many sets of parameters $\boldsymbol{\theta}=(a,b,c,d,k,G_{5})$ to select in preparing training dataset for latter emulator construction, should be carefully chosen to balance the computational efficiency against statistical robustness, with the aim to minimize computational expense while ensuring sufficient coverage of the parameter space and accuracy in Bayesian posterior estimation.
For this purpose, we select $d$ design points, which are structured into a $d\times 6$ design matrix $\mathbf{\Theta}=(\boldsymbol{\theta}_{1},\boldsymbol{\theta}_{2},...,\boldsymbol{\theta}_{d})$, where each row corresponds to a unique parameter set. The allowable range for each parameter is defined to be physically broad enough to encompass all plausible configurations while avoiding artificial truncation that could bias the inference.

Next, we emply Latin hypercube sampling (LHS)\cite{c42ec141-810d-3108-8740-320eb5d0f4b6, MORRIS1995381} to distribute the design points uniformly across the 6-dimensional parameter space. LHS is a space-filling statistical method optimized for generating representative samples in high-dimensional parameter spaces. Unlike traditional random sampling, LHS partitions each parameter's range into $d$ equal subintervals and draws one sample per subinterval, encuring uniform marginal coverage across all dimensions.  
The advantage of this stratified sampling method is that it ensures a more uniform distribution of data points, avoiding clustering or missing certain areas. Even with limited computational resources and fewer sample points, it can efficiently cover the entire parameter space. This characteristic is particularly suitable for subsequent modeling (such as emulators like Gaussian processes), where uniformly distributed data can enable more accurate interpolation by the model in high-dimensional parameter spaces, thereby improving the precision of Bayesian inference.

For the 6-dimensional parameter space of the EMD model ($\boldsymbol{\theta}=(a,b,c,d,k,G_{5})$) specifically, we implemented LHS to generate 300 design points, organized into a $300\times 6$ design matrix
$\mathbf{\Theta}=(\boldsymbol{\theta_{1}},\boldsymbol{\theta_{2}},...,\boldsymbol{\theta_{300}})$. While prior studies\cite{article} suggest that $\sim 10$ design points typically suffice for acceptable computational accuracy, we prioritized enhanced emulator performance by selecting 300 points -- a balance between computational feasibility and statistical fidelity. Each parameter set $\theta_i$ was propagated through the EMD model to compute the corresponding observables $S_{BH}$, $\chi_{2}^{B}$ and $C_{s}^{2}$. The resulting 300 output sets were post-processed (see next subsetion) to enable comparison with Lattice QCD constraints.

\subsection{Postprocessing model output}

\begin{figure*}[tbp!]
     \centering
     \includegraphics[width=1.0\textwidth]{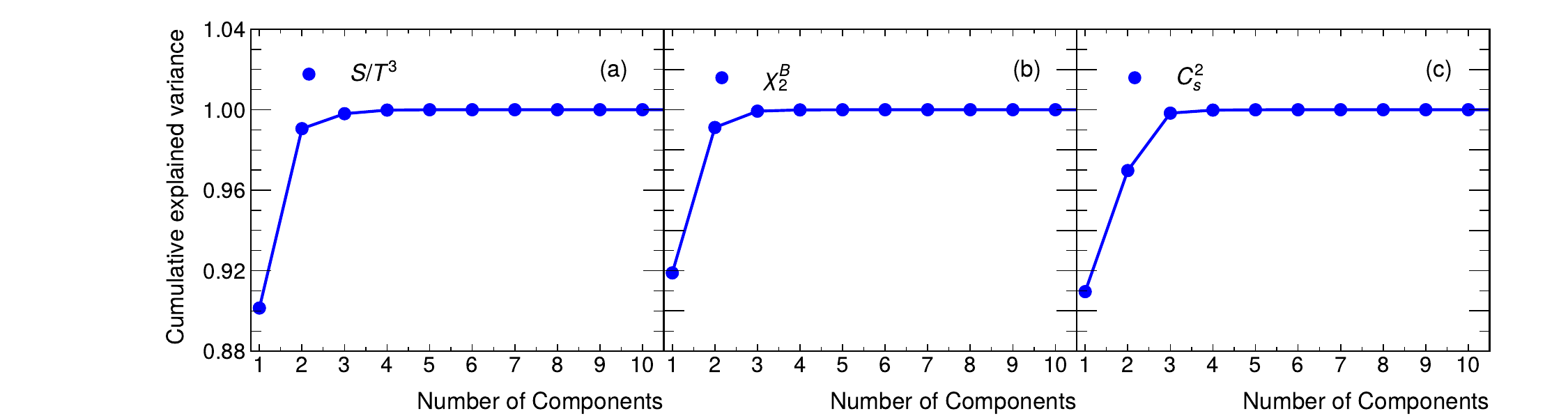}
     \caption{Cumulative explained variance versus principla component (PC) count for (a) $S_{BH}$, (b) $\chi_{2}^{B}$ and (c) $C_{s}^{2}$. The first 3 PCs capture $99\%$ of the variance in all cases, while the first 6 PCs are retained for conservative emulation.}
     \label{fig2}
\end{figure*}

Taking as input with the sampled parameter sets $\mathbf{\Theta}=(\boldsymbol{\theta}_{1},\boldsymbol{\theta}_{2},...,\boldsymbol{\theta}_{300})$, the EMD model generates 300 output vectors $\mathbf{y}=(y_{1},y_{2},...,y_{m})$, where $m$ corresponds to the number of temperature-dependent data points for the lattice observables $S_{BH}$ (55 points), $\chi_{2}^{B}$ (19 points) and $C_{s}^{2}$ (55 points) as defined in Equations (23), (30), and (28). 
While constructing independent Gaussian process (GP) emulators for each output variable (see next subsection) is straightforward, direct usage of these outputs in doing so scales poorly with $m$ and neglects inter-output correlations. To address these limitations, we postprocess them by applying
Principal Component Analysis (PCA) \cite{10.1162/089976699300016728}, which reduces the output dimensionality and decorrelate the data while preserving covariant structure. 

PCA projects correlated high-dimensionaly data into a lower-dimensional subspace spanned by orthogonal principal components (PCs) ordered by their explained variance.
For our case, the outputs are organized into a $300\times m$ matrix $Y$, where rows represent design points and columns correspond to observables at discrete temperatures. Prior to PCA, $Y$ is standardized by centering each column (subtracting the mean) and scaling to unit variance, ensuring equal weighting across temperature points.

The standardized data is decomposed via singular value decomposition (SVD) to obtain the orthonormal transformation matrix $V$, whose columns $v_j$ define the PC directions. The projected data in PC space is given by $Z = Y V$, where $Z$ is a $300\times m$ matrix. Each row of $Z$ corresponds to a design point, while columns represent PCs ranked by descending variance contribution, The first PC captures the dominant covariance pattern in $Y$, with subsequent PCs encoding orthogonal modes of decreasing significance.

For correlated physical observables, the leading PCs typically encapsulate the majority of system variance. Fig. \ref{fig2} illustrates the cumulative explained variance versus the number of PCs retained for $S_{BH}$, $\chi_{2}^{B}$ and $C_{s}^{2}$. While $99\%$ of the variance is captured by the first three PCs across all observables, we conservatively retain the 6 PCs to ensure robust emulation of subtle features. This reduces the effective dimensionality from $m = 55, 19, 55$ to 6, drastically lowing computational costs without sacrificing predictive accuracy.

The PCA workflow was implemented using the scikit-learn library in Python, leveraging its optimized numerical routines for efficient SVD computation and component projection. The retained PCs, $z_1, z_2, ..., z_6$, serve as uncorrelated targets for subsequent Gaussian process emulation, enabling efficient Bayesian inference while preserving inter-observable correlations.

\subsection{Gaussian Processes Emulator}
To reduce computational costs while preserving predictive accuracy later in the extensive Bayesian exploration for the posterior distribution, we employ a Gaussian process (GP) emulator\cite{williams2006gaussian} as a surrogate model for the full EMD model calculation.
A GP is a probabilistic model defines a distribution over functions where any finite set of function evaluations follows a multivariate Gaussian distribution. Formally, for input parameters $\boldsymbol{\theta}\in \mathcal{R}^6$ (with design matrix $\mathbf{\Theta}=(\boldsymbol{\theta}_{1},\boldsymbol{\theta}_{2},...,\boldsymbol{\theta}_{300})$) and outputs $\mathbf{Y}=(\mathbf{y}_{1},\mathbf{y}_{2},...,\mathbf{y}_{300})$, the GP is fully specified by: (1), A \textbf{mean function} $\mu(\boldsymbol{\theta})$: Typically set to zero after centering the data ($Y\leftarrow Y - \text{mean}(Y)$), simplifying prior assumptions; and (2), A \textbf{covariance kernel} $\sigma(\boldsymbol{\theta}, \boldsymbol{\theta}')$: Encodes correlations between outputs at different input points. 
With these two functions, the Gaussian Process provides a flexible framework for modeling complex dependencies and uncertainty in data.

The squared-exponential kernel is adopted for its smoothness and interpretability:
\begin{align}
\sigma(\boldsymbol{\theta}, \boldsymbol{\theta}') = \exp\left(-\frac{|\boldsymbol{\theta} - \boldsymbol{\theta}'|^2}{2\ell^2}\right),
\label{eq:SE_kernel }
\end{align}
where $l>0$ is the characteristic length scale governing correlation decay rate between input points: Nearby points ($||\boldsymbol{\theta}-\boldsymbol{\theta}'||\ll l$) are strongly correlated, while distant points ($||\boldsymbol{\theta} - \boldsymbol{\theta}'||\gg l$) become independent.

For $n=300$ training points, the outputs $Y$ jointly follow a multivariate normal distribution:
\begin{align}
\begin{split}
\mathbf{Y} &\sim \mathcal{N}\left(\mathbf{0},\mathbf{K}\right),\\ 
\mathbf{K} &=
\begin{pmatrix}
\sigma(\boldsymbol{\theta}_1, \boldsymbol{\theta}_1) & \sigma(\boldsymbol{\theta}_1, \boldsymbol{\theta}_2) & \cdots & \sigma(\boldsymbol{\theta}_1, \boldsymbol{\theta}_{300}) \\
\vdots & \vdots & \ddots & \vdots \\
\sigma(\boldsymbol{\theta}_{300}, \boldsymbol{\theta} _1) & \sigma(\boldsymbol{\theta}_{300}, \boldsymbol{\theta}_2) & \cdots & \sigma(\boldsymbol{\theta}_{300}, \boldsymbol{\theta}_{300})
\end{pmatrix},
\label{eq:GP_prior }
\end{split}
\end{align}
where $K$ is the $300\times 300$ covariance matrix.

Then, when given test inputs $\mathbf{\Theta}_{*}$, GP predict the distribution for its outputs $\mathbf{Y}_{*}$ is Gaussian as following:
\begin{align}
\mathbf{Y}_{*} |\mathbf{\Theta_{*}}, \mathbf{\Theta}, \mathbf{Y} &\sim \mathcal{N}\left(\boldsymbol{\mu}_{*}, \mathbf{\Sigma}*\right),
\label{eq:GP_posterior }
\end{align}
\begin{align}
\boldsymbol{\mu}_{*} &= \mathbf{K}{*} \mathbf{K}^{-1} \mathbf{Y},
\label{eq:GP_mean } \
\end{align}
\begin{align}
\mathbf{\Sigma}* &= \mathbf{K}{**} -  \mathbf{K}_{*} \mathbf{K}^{-1} \mathbf{K}^\mathrm{T}_{*},
\label{eq:GP_cov }
\end{align}
where $\mathbf{K}_{*}=\sigma(\mathbf{\Theta_{*}},\mathbf{\Theta})$ and $\mathbf{K}_{**}=\sigma(\mathbf{\Theta_{*}},\mathbf{\Theta_{*}})$.

The EMD model is treated as a latent function $f(\boldsymbol{\theta})$ with input $\boldsymbol{\theta}=(a,b,c,d,k,G_5)$, where $f$ maps them to observables (more specifically, the first 6 principle components of) $S_{BH}$, $\chi_2^B$, and $C^2_s$. The GP emulator training proceeds as follows: (1), Center outputs $Y$ by substracting the empirical mean; (2), Maximize the marginal likelihood to determin $l$ for the kernel; (3), Use Equations~(33) - (35) to predict $Y_{*}$ at new $\mathbf{\Theta_{*}}$.

This framework enables efficient interpolation across the 6D parameter space while quantifying predictive uncertainty—critical for Bayesian parameter inference.

\subsection{Bayesian extraction of the model parameters}

We are now able to calibrate our EMD model with lattice data (also called evidence in Bayesian inference), specifially, to derive numerical evaluations of the model parameters while simultaneously evaluating their uncertainties. This is essentially an inverse problem, the primary objective of which is to infer unknown model input parameters or physical factors using the available output data\cite{Pang:2016vdc,OmanaKuttan:2020btb,Jiang:2021gsw,Aarts:2025gyp}.  
Within the Bayesian inference framework, the estimation of model parameters is derived from their posterior distribution, which combines prior information with likelihood function constructed from observational data to deliver more reliable parameter estimation results,
\begin{align}
\mathit{P}(\boldsymbol{\theta} | data)\propto\mathit{P}(data |\boldsymbol{\theta})\mathit{P}(\boldsymbol{\theta})
\end{align}
In this context, $\boldsymbol{\theta}$ represents the parameters of the model, while data encompasses the evidence we choose from lattice simulation. 
This relationship, known as Bayes' theorem, expresses the posterior distribution on the left-hand side, which describes the probability distribution of the parameters conditioned on the observed data. 
On the right-hand side, $\mathit{P}(data|\boldsymbol{\theta})$ denotes the likelihood function, indicating the chance that the chosen evidence (data) to be observed 
under particular parameter values. Meanwhile, $\mathit{P}(\boldsymbol{\theta})$ corresponds to the prior distribution, which encapsulates our initial beliefs or assumptions about the parameters before incorporating the observed data.
 
In the following, we will elaborate on the definitions of the prior distribution and likelihood function, as well as how to efficiently sample from the posterior distribution using the MCMC method. 

First, the prior distribution $\mathit{P}(\boldsymbol{\theta})$ captures initial assumptions about the parameters before observing any data. When no specific knowledge is available, a uniform prior is a widely used option, assigning a constant value to $\mathit{P}(\boldsymbol{\theta})$ . In this work, each parameter is assumed to be uniformly distributed within a limited design range, thus the Gaussian process emulator's predictions are valid only within these bounds. Accordingly, the prior can be defined as a constant within the design region and zero elsewhere, providing both a practical constraint and a streamlined representation of prior knowledge.
\begin{align}
P(\mathbf{\boldsymbol{\theta}}) \propto 
\begin{cases} 
1 & \text{if } \min(\boldsymbol{\theta}_i) \leq \boldsymbol{\theta}_i \leq \max(\boldsymbol{\theta}_i) \text{ for all } i, \\
0 & \text{else.}
\end{cases}
\end{align}
Setting the prior to zero outside the design region is a strict and cautious assumption, effectively ruling out any possibility of the true parameter values lying beyond the defined limits. However, this rigidity risks overlooking valid parameter spaces, potentially undermining the model's flexibility and predictive power. A more sensible approach is to define broader design ranges, which allow for greater uncertainty in parameter estimation. This ensures the model remains versatile and captures a wider spectrum of plausible parameter values, enhancing both its robustness and practical relevance.

\begin{figure}[h]
    \centering
    \includegraphics[width=0.32\textwidth]{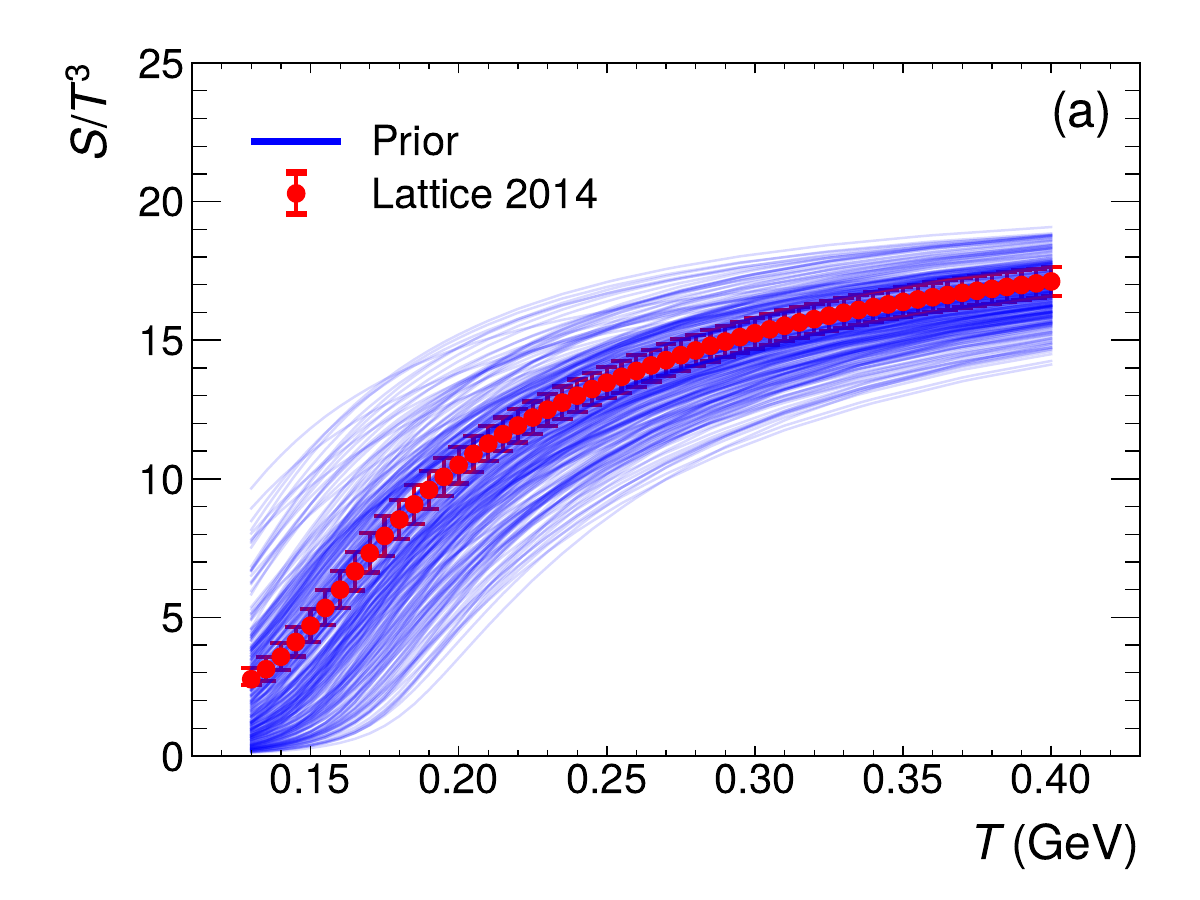}
    \includegraphics[width=0.32\textwidth]{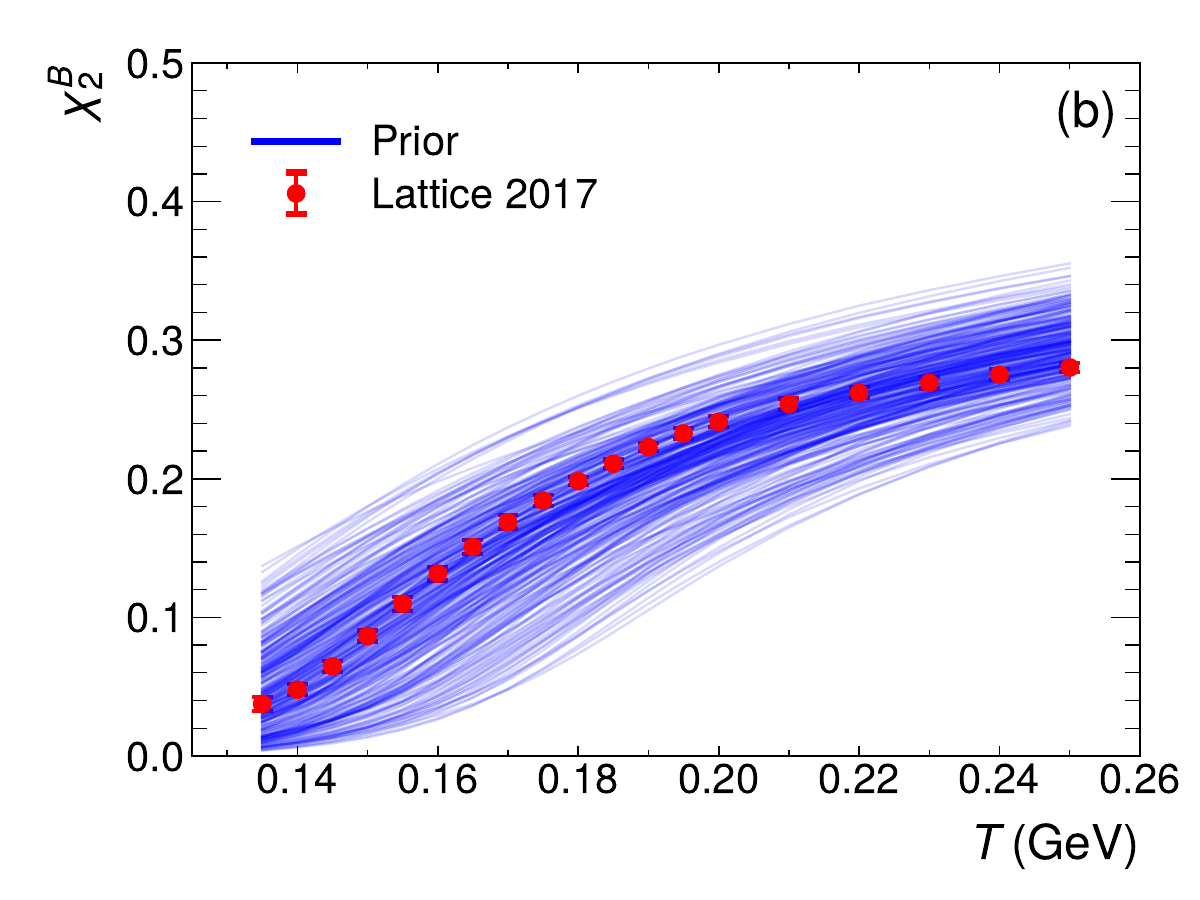}
    \includegraphics[width=0.32\textwidth]{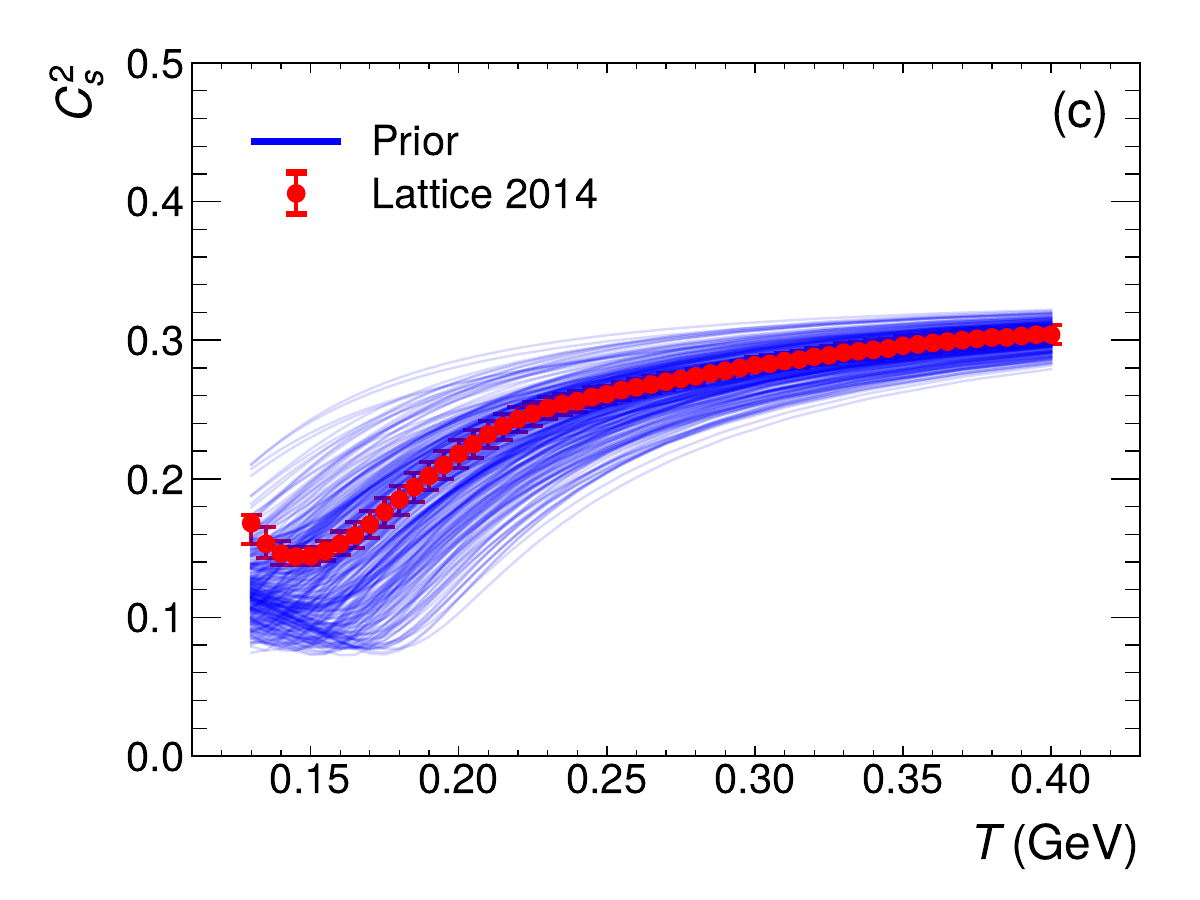}
     \caption{(a) The prior distribution of entropy as a function of temperature. (b) The prior distribution of baryon number susceptibility as a function of temperature. (c) The prior distribution of the square of the speed of sound as a function of temperature. The lattice results from Ref \cite{HotQCD:2014kol,Bazavov:2017dus}.}
     \label{fig3}
\end{figure}

In our study, the prior ranges for the model parameters $\boldsymbol{\theta}(a,b,c,d,k,G_{5})$ are comprehensively listed in Table \ref{table1}. By applying these prior ranges to the EMD model, we derived theoretical predictions for $S_{BH}$, $\chi_{2}^{B}$ and $C_{s}^{2}$ , which are visually presented in Fig. \ref{fig3}.

\begin{table}[h]
\centering
\begin{tabular}{|c|c|c|}
  \hline
  \multicolumn{3}{|c|}{Prior} \\ \hline
  Parameter & min & max \\ \hline
  $a$ & 0.110 & 0.310 \\ \hline
  $b$ & 0.005 & 0.031 \\ \hline
  $c$ & -0.280 & -0.205 \\ \hline
  $d$ & -0.240 & -0.110 \\ \hline
  $k$ & -0.910 & -0.770 \\ \hline
  $G_{5}$ & 0.375 & 0.430 \\ \hline
\end{tabular}
\caption{Parameter ranges for prior distribution in the EMD model.} 
\label{table1}
\end{table}

Next, the likelihood function of the parameters given the observed data, denoted as $\mathit{P}(data|\boldsymbol{\theta})$, quantifies the probability of observing specific data under the model with a given parameter $\boldsymbol{\theta}$, which due to the central limit theorem can be expressed as Gaussian form with the chi-square as the exponent: 
\begin{align}\label{PPP}
\mathit{P}(data |\boldsymbol{\theta})=\underset{i}{\prod}\frac{1}{\sqrt{2\pi}\sigma_{i}}e^{-\frac{[y_{i}(\boldsymbol{\theta})-y_{i}^{\text{lattice}}]^{2}}{2\sigma_{i}^{2}}}
\end{align}
where $y_{i}(\boldsymbol{\theta})$ and $y_i^{\text{lattice}}$ denote our GP emulator predicted observables (specifially, the first 6 PCs) and those from lattice simulation. Note that here $y_{i}^{\text{lattice}}$ is obtained by applying the same transformation matrix $V$ derived from the PCA as specified in subsection B on the lattice simulated $S_{BH}$, $\chi_{2}^{B}$ and $C_{s}^{2}$.
In our calculations,  the error bars are self-consistently incorporated into the posterior computation within the Bayesian inference framework, specifically reflected in the $\sigma^{2}$ of the exponential part of our likelihood (Eq. \ref{PPP}). The error $\sigma$ includes the error from the lattice data and the error from the Gaussian process emulator, $\sigma^{2} = \sigma_{\mathrm{lattice}}^{2} + \sigma_{\mathrm{Gaussian \;emulator}}^{2}$.

In our calculation, we first compute independent likelihood functions for $S_{BH}$, $\chi_{2}^{B}$ and $C_{s}^{2}$, then construct a joint likelihood function as the product of indidual likelihoods:
\begin{align}
\mathit{P}(data |\boldsymbol{\theta})=\mathit{P}(data_{S_{BH}} |\boldsymbol{\theta})\mathit{P}(data_{\chi_{2}^{B}} |\boldsymbol{\theta})\mathit{P}(data_{C_{s}^{2}} |\boldsymbol{\theta})
\end{align}
This multiplicative approach systematically integrates constraints from all observables, ensuring a holistic assessment of parameter compatibility.

Finally, to sample the posterior distribution $\mathit{P}(\boldsymbol{\theta} |data)$, we employ the Metropolis-Hastings algorithm \cite{Metropolis:1953am, Hastings:1970aa}, a straightforward and commonly used Markov Chain Monte Carlo (MCMC) sampling technique in Bayesian inference for handling high-dimensional parameter spaces and complex distributions. It iteratively samples by randomly proposing a candidate $\boldsymbol{\theta}^{\prime}$ from the current position $\boldsymbol{\theta}_i$, determining acceptance based on their posterior probabilities. If accepted, the next position is updated to $\boldsymbol{\theta}_{(i+1)}=\boldsymbol{\theta}^{\prime}$; otherwise, it remains at $\boldsymbol{\theta}_{(i+1)}=\boldsymbol{\theta}_{(i)}$. Repeating this process produces a sequence $\{\boldsymbol{\theta}_{(1)},\boldsymbol{\theta}_{(2)},...,\boldsymbol{\theta}_{(n)}\}$ that eventually converges to the target posterior distribution. After discarding initial ``burn-in'' samples (typically $10 - 20\%$ of the chain) to mitigate transient correlation effects from starting values of the parameters, the remaining chain approximates the target posterior.

To ensure better numerical stability and efficiency for high-dimensional probability calculations, we work with the log-posterior:
\begin{align}
\mathrm{log}P(\boldsymbol{\theta}| data)=\mathrm{log}P(data|\boldsymbol{\theta})+\mathrm{log}P(\boldsymbol{\theta})+\mathrm{const}
\end{align}
where additive constants after taking logarithm of the distribution are irrelevant for MCMC, which depends only on probability ratios. 
Thus, terms associated with those additive constants can be disregarded. Specifically, the logarithmic expressions for the prior Equation (37) and the uniform likelihood Equation (39) are as follows:
\begin{align}
\mathrm{log}P(\boldsymbol{\theta}) \propto 
\begin{cases} 
0 & \text{if } \min(\boldsymbol{\theta}_i) \leq \boldsymbol{\theta}_i \leq \max(\boldsymbol{\theta}_i) \text{ for all } i, \\
-\infty & \text{else.}
\end{cases}
\end{align}
\begin{align}
\begin{split}
\mathrm{log}P(data|\boldsymbol{\theta})&=-\frac{1}{2}\mathbf{y}^{T}\Sigma^{-1}\mathbf{y}-\frac{1}{2}\mathrm{log}\lvert\Sigma\rvert-\frac{1}{2}\mathrm{log}2\pi,\\
\mathbf{y}&=y(\boldsymbol{\theta})-y^{\mathrm{lattice}}
\end{split}
\end{align}

The initial ``burn-in'' phase allows the chain to transition from arbitrary starting points to regions of high posterior density. The required burn-in length depends on the posterior's complexity and the proposal distribution's efficiency. In practice, 
we employed 120 walkers, which were randomly initialized within the parameter space. An initial 10000 steps were run as a burn-in phase to ensure that the chain adequately converged to the target posterior distribution. After the burn-in phase, an additional 10000 steps were conducted to generate posterior samples. This sample size is sufficient for constructing smooth histogram visualizations; however, it may be excessive for tasks such as calculating medians or other summary statistics. It is worth emphasizing that the total number of samples is determined by the product of the number of walkers and the number of steps.

Once the MCMC sampling is completed, the marginal distributions for each parameter within $\boldsymbol{\theta} = (a,b,c,d,k,G_{5})$ can be derived. A marginal distribution is obtained from the posterior distribution by isolating a specific parameter or a group of parameters of interest, while integrating out the remaining parameters. For example, in our model, the marginal distribution of parameter $d$ can be determined by eliminating the contributions of all other parameters through integration.
\begin{align}
P(d| data)=\int da db dc dk dG_5P(\boldsymbol{\theta}| data)
\end{align}

\begin{figure*}[tbp!]
     \centering
     \includegraphics[width=0.8\textwidth]{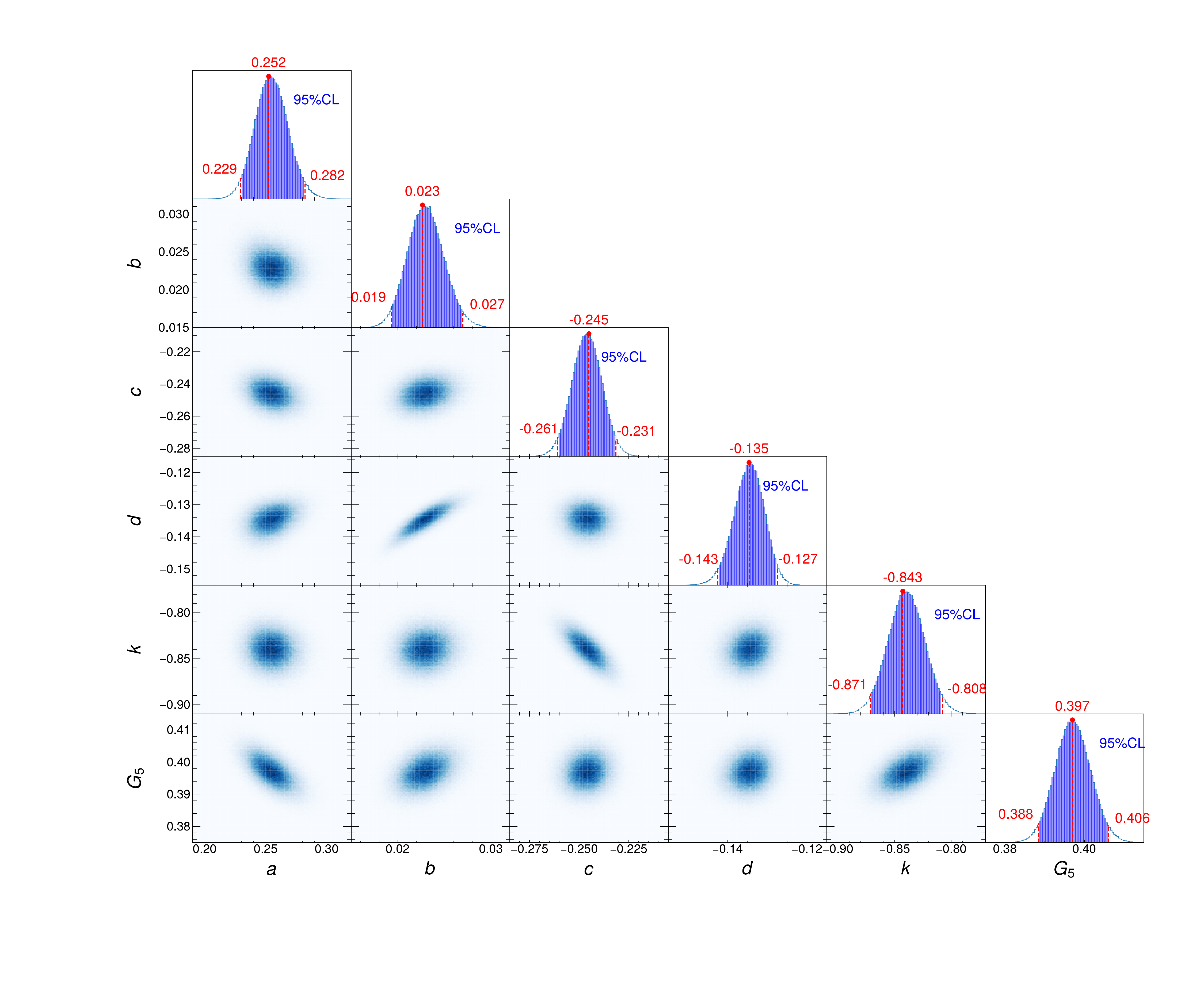}
     \caption{Diagonal: The marginal posterior distributions of the model parameters, along with the MAP values and the 95\% confidence levels for each parameter. Lower triangle: The joint distribution between each pair of parameters.}
     \label{fig4}
\end{figure*}

For the joint distribution of two parameters, such as $a$ and $d$, it can be calculated as:
\begin{align}
P(a,d| data)=\int db dc dk dG_5P(\boldsymbol{\theta}| data)
\end{align}

\begin{table}[h] 
\centering 
\begin{tabular}{|c|c|c|c|} 
  \hline
  \multicolumn{4}{|c|}{Posterior  95\% CL} \\ \hline
  Parameter & min & max & MAP \\ \hline
  $a$ & 0.229 & 0.282 & 0.252 \\ \hline
  $b$ & 0.019 & 0.027 & 0.023 \\ \hline
  $c$ & -0.261 & -0.231 & -0.245 \\ \hline
  $d$ & -0.143 & -0.127 & -0.135 \\ \hline
  $k$ & -0.871 & -0.808 & -0.843 \\ \hline
  $G_{5}$ & 0.388 & 0.406 & 0.397 \\ \hline
\end{tabular}
\caption{Parameter estimates with posterior 95\% confidence levels and MAP values for parameters of the EMD model.} 
\label{table2}
\end{table}

With the evidence (data) chosen as $S/T^{3}, \chi^B_2$ and $C^2_s$ from the lattice QCD\cite{HotQCD:2014kol, Bazavov:2017dus} into our Bayesian analysis, we add $C^2_s$ to better highlight the phase transition region.
Table \ref{table2} summarizes our inferred parameter constraints, reporting the $95\%$ credible intervals and maximum a posteriori (MAP) estimates derived from the MCMC.
Fig. \ref{fig4} visualizes the posterior distributions using a corner plot (or triangle plot), a standard tool in Bayesian analysis for high-dimensional parameter spaces. 
It consists of a diagonal subplots and off-diagonal subplots. 
The diagonal subplots show the marginal distributions of individual parameters, i.e., the marginal probability density of each parameter after integrating out other parameters, and we took the 95\% confidence levels and identified the maximum posterior value. The off-diagonal subplots display the joint distributions between pairs of parameters, visualized as 2D density estimates, revealing correlations or degeneracies between parameters, such as trade-offs in how combinations of different parameters accommodate observational constraints. It's evident that the used data from lattice can well calibrate each parameter in our model with clear peak structure, and certain correlations between several parameters like $b$ and $d$ are also shown.

\section{Thermodynamics in the bayesian holographic QCD model} \label{sec4}

This section systematically evaluates the thermodynamic predictions of our Bayesian-calibrated holographic model. We first analyze the impact of evidence selection by comparing results from two cases in subsection A: (1), case 1: with $S/T^3$ and $\chi^B_2$ as evidence; and (2) case 2: with $S/T^3$, $\chi^B_2$ and $C^2_s$ jointly as evidence. Subsequently, in subsection B we confront our results from case 2 against lattice QCD \cite{HotQCD:2014kol, Bazavov:2017dus}, the Hadron Resonance Gas (HRG) model \cite{Vovchenko:2019pjl}, and the Hard Thermal Loop (HTL) model \cite{Haque:2014rua}.


\begin{figure*}[tbp!]
    \centering
    \begin{minipage}{0.33\textwidth}
        \includegraphics[width=\textwidth]{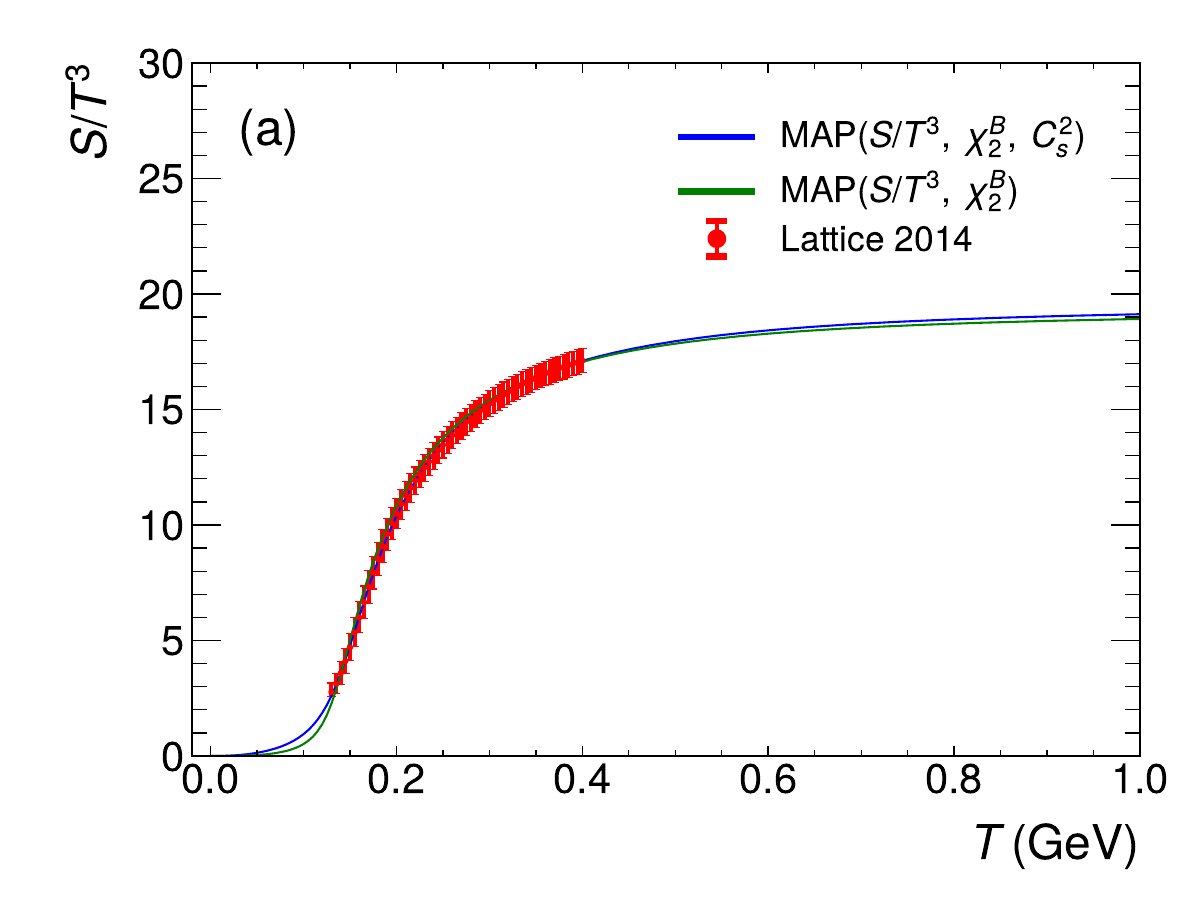}
    \end{minipage}
    \begin{minipage}{0.33\textwidth}
        \includegraphics[width=\textwidth]{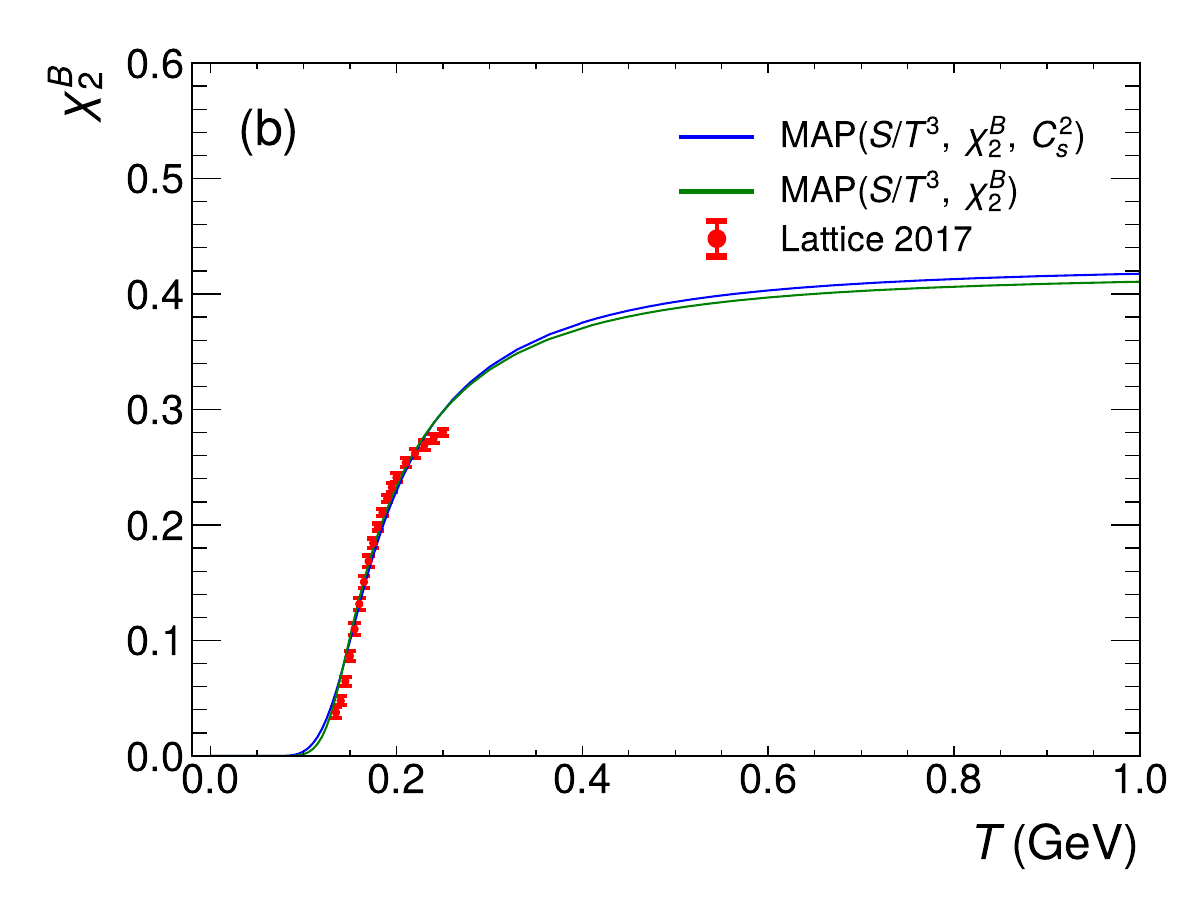}
    \end{minipage}
    \begin{minipage}{0.33\textwidth}
        \includegraphics[width=\textwidth]{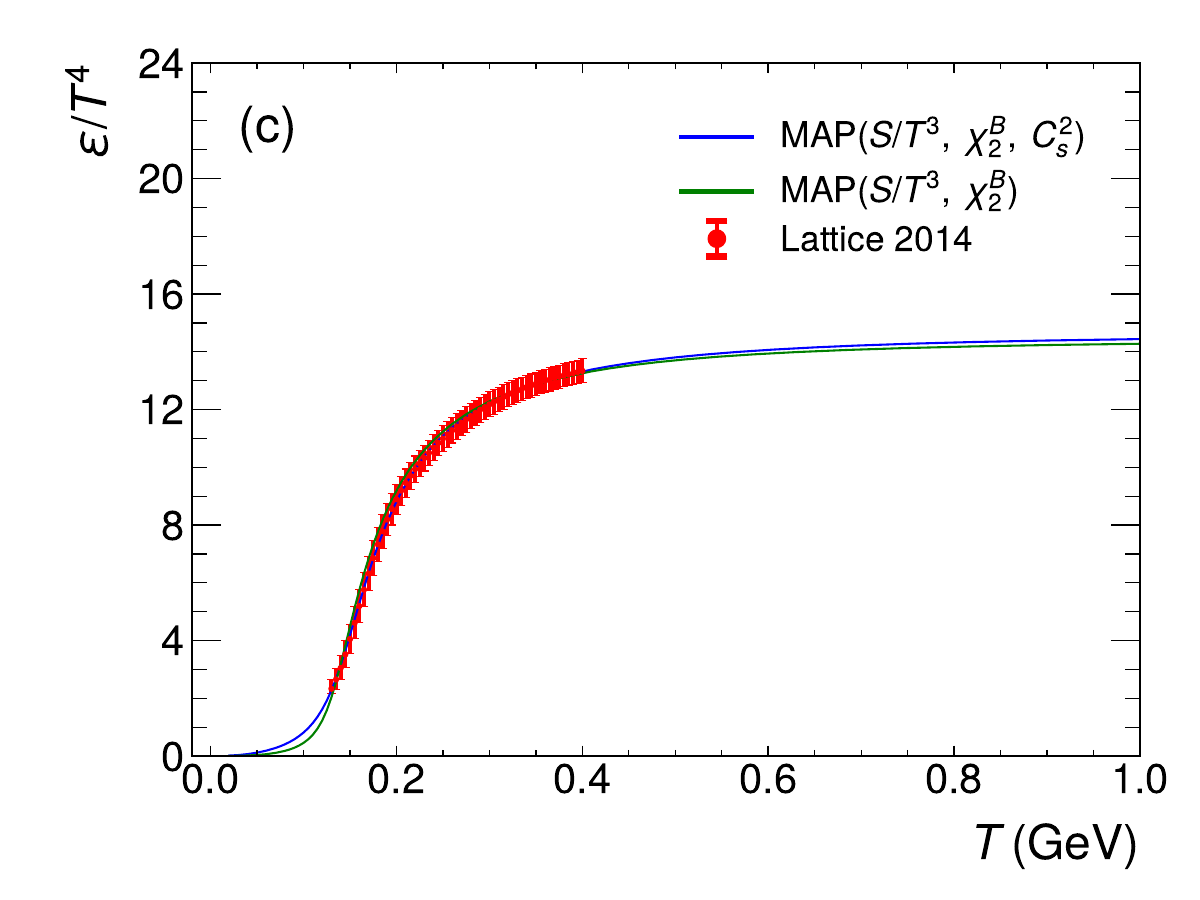}
    \end{minipage}
    \begin{minipage}{0.33\textwidth}
        \includegraphics[width=\textwidth]{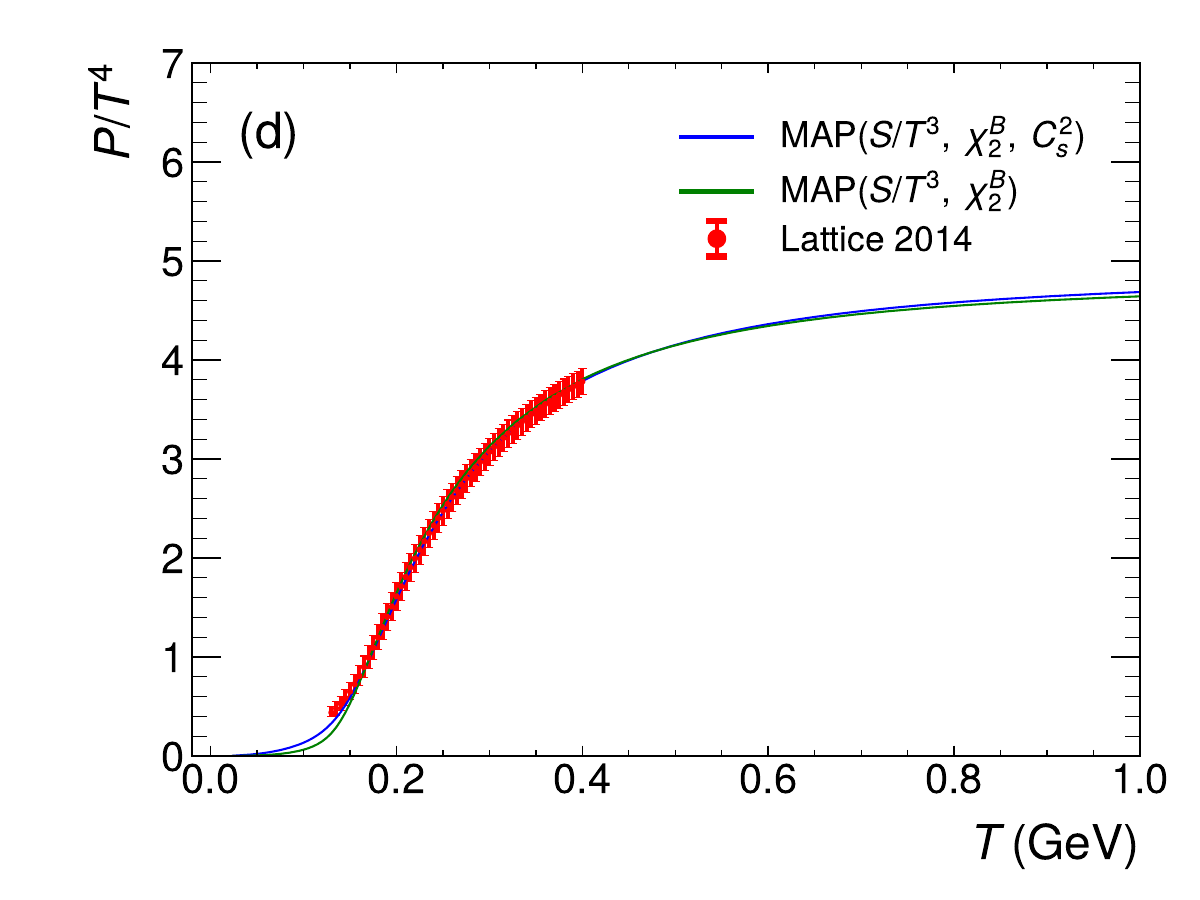}
    \end{minipage}
    \begin{minipage}{0.33\textwidth}
        \includegraphics[width=\textwidth]{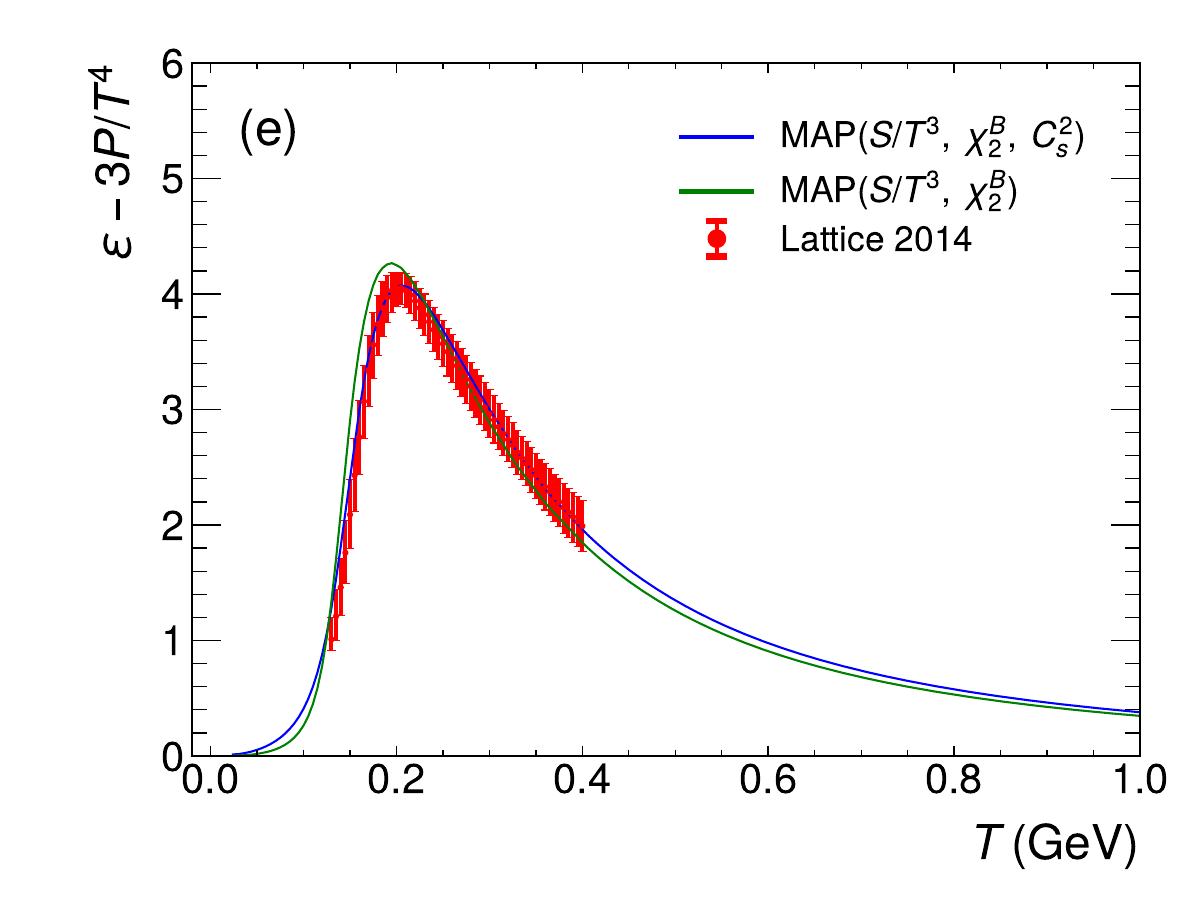}
    \end{minipage}
    \begin{minipage}{0.33\textwidth}
        \includegraphics[width=\textwidth]{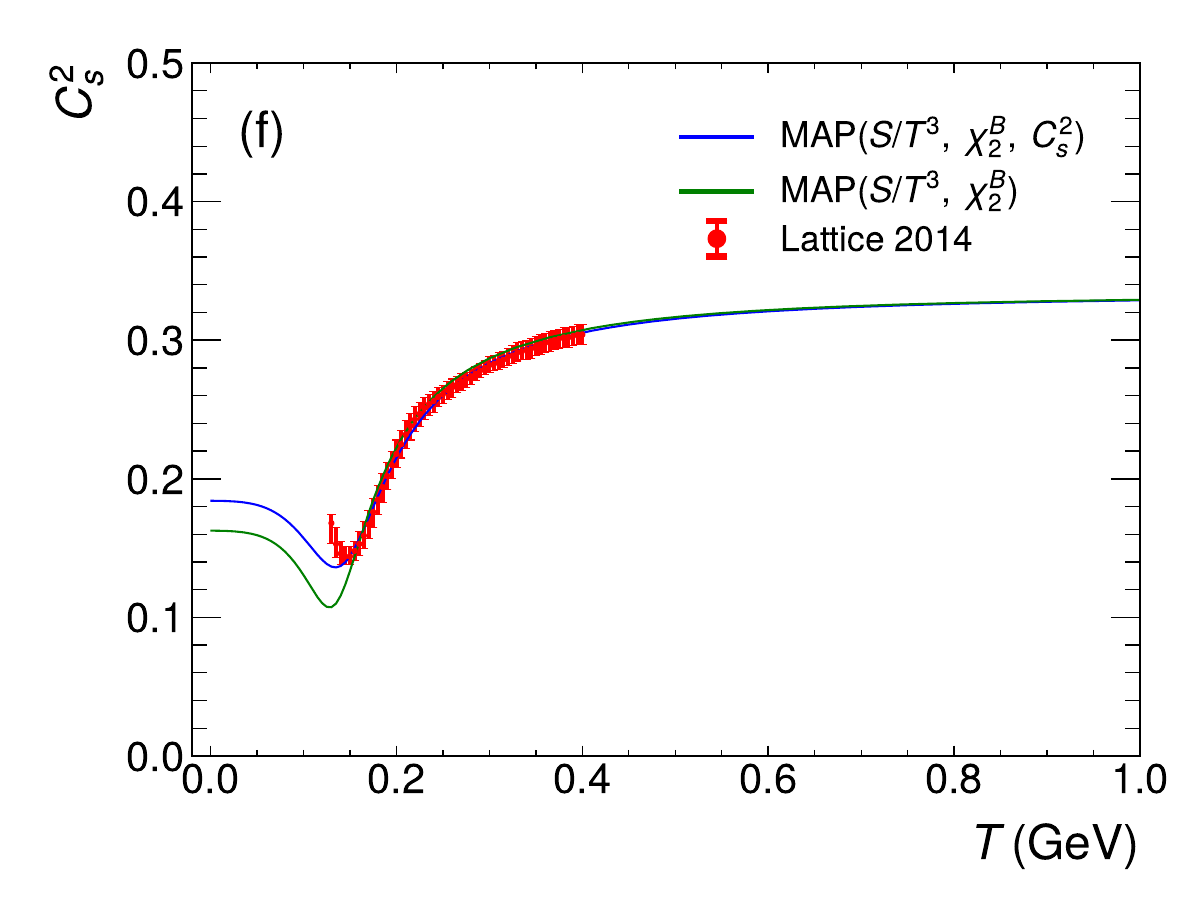}
    \end{minipage}
    \begin{minipage}{0.33\textwidth}
        \includegraphics[width=\textwidth]{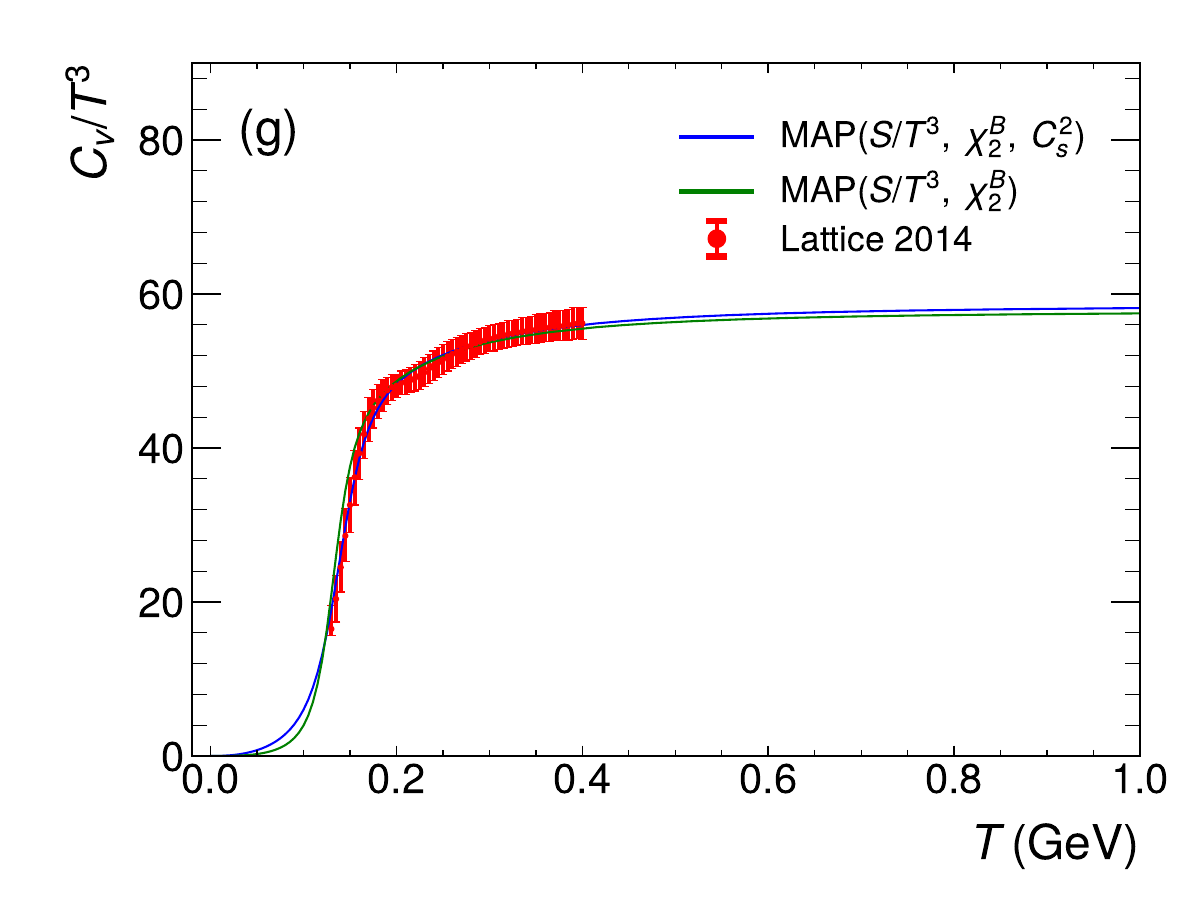}
    \end{minipage}
    \begin{minipage}{0.33\textwidth}
        \includegraphics[width=\textwidth]{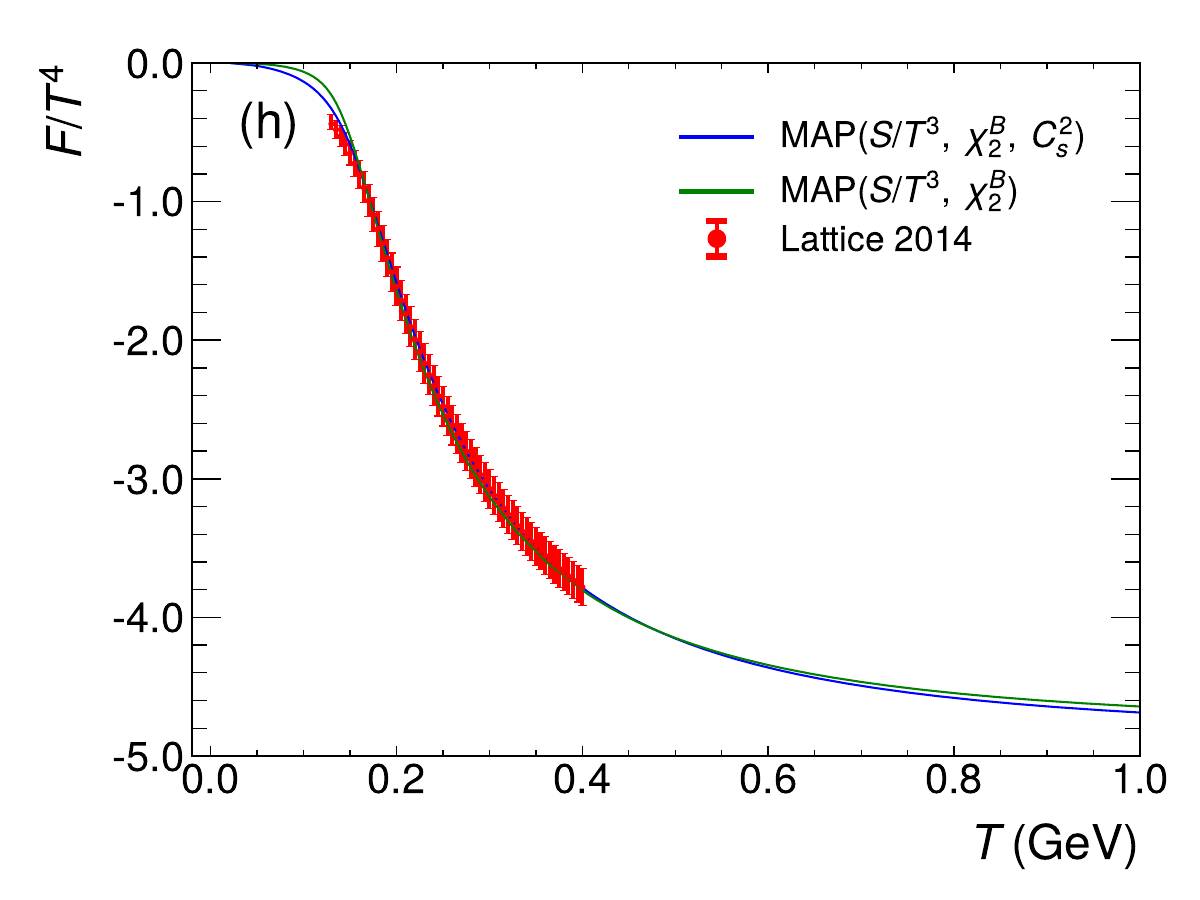}
    \end{minipage}
    \begin{minipage}{0.33\textwidth}
        \includegraphics[width=\textwidth]{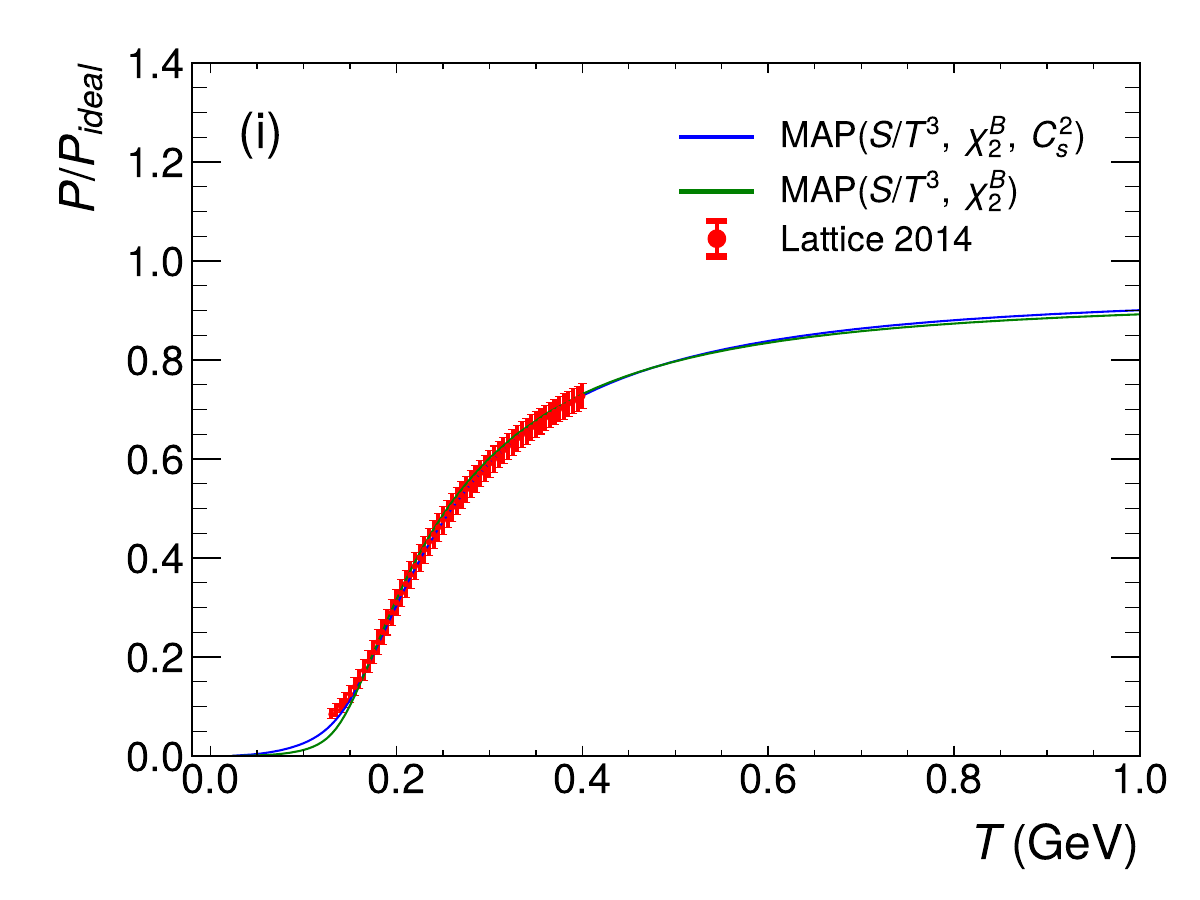}
    \end{minipage}
    \begin{minipage}{0.33\textwidth}
        \includegraphics[width=\textwidth]{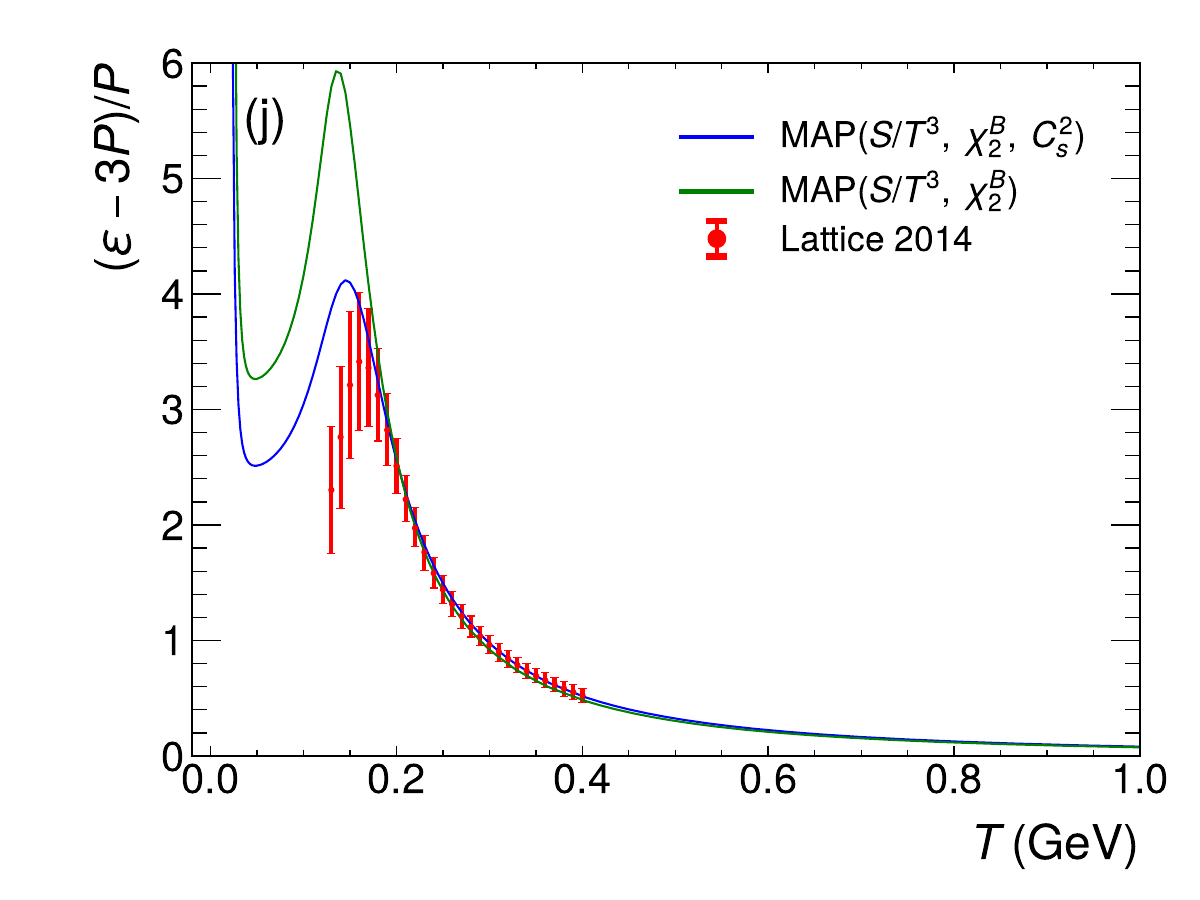}
    \end{minipage}
    \caption{Using Bayesian inference to compare the posterior thermodynamic results under case 1 and case 2, the green curves represents the calculated results of the MAP value in case 1, while the blue curves represents the calculated results of the MAP value in case 2. Subfigures a - j correspond to the entropy, susceptibility, energy, pressure, trace anomaly, the square of the speed of sound, the specific heat, free energy, the ratio of pressure to the ideal gas limit pressure and the ratio of the trace anomaly to pressure as functions of temperature, respectively. The lattice results are taken from Refs. \cite{HotQCD:2014kol, Bazavov:2017dus}.}
    \label{fig5}
\end{figure*}


\begin{figure*}[tbp!]
    \centering
    \begin{minipage}{0.33\textwidth}
        \includegraphics[width=\textwidth]{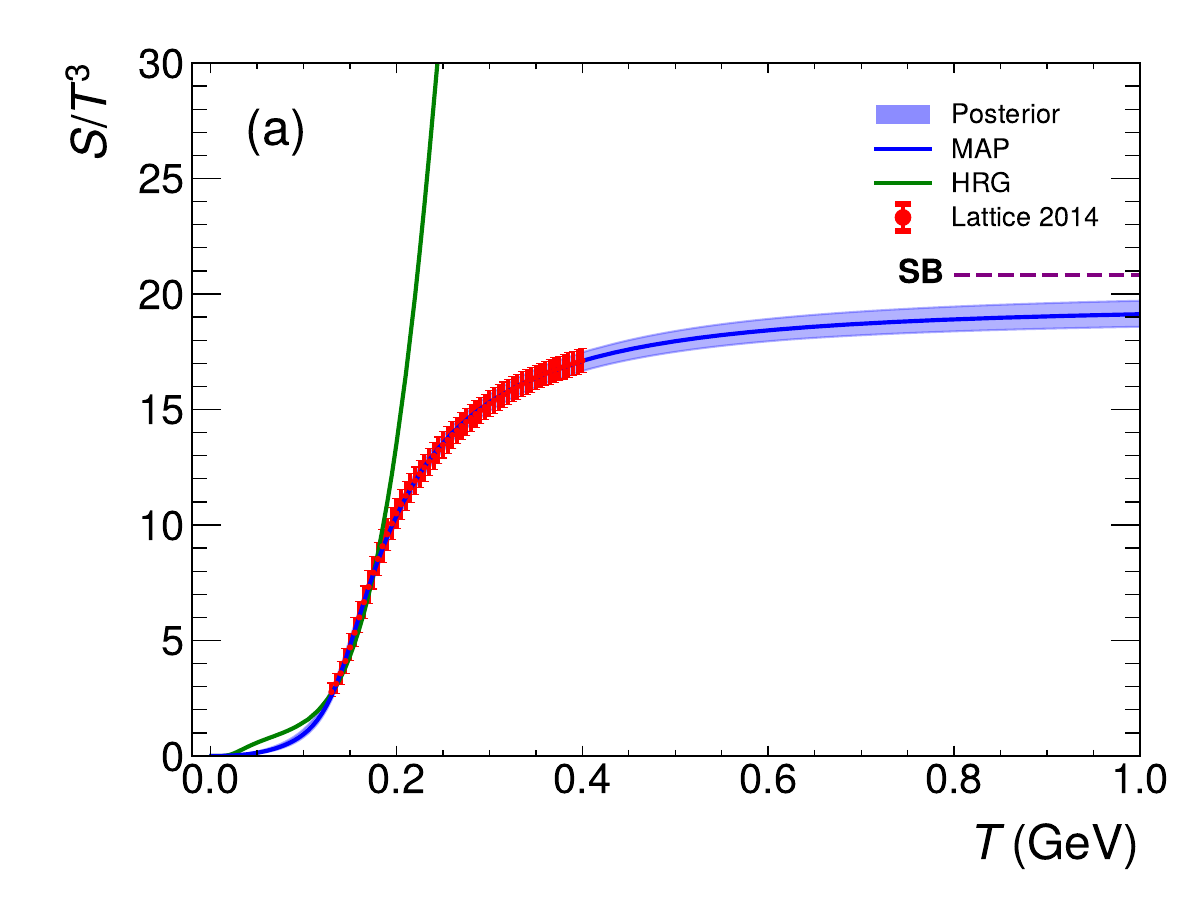}
    \end{minipage}
    \begin{minipage}{0.33\textwidth}
        \includegraphics[width=\textwidth]{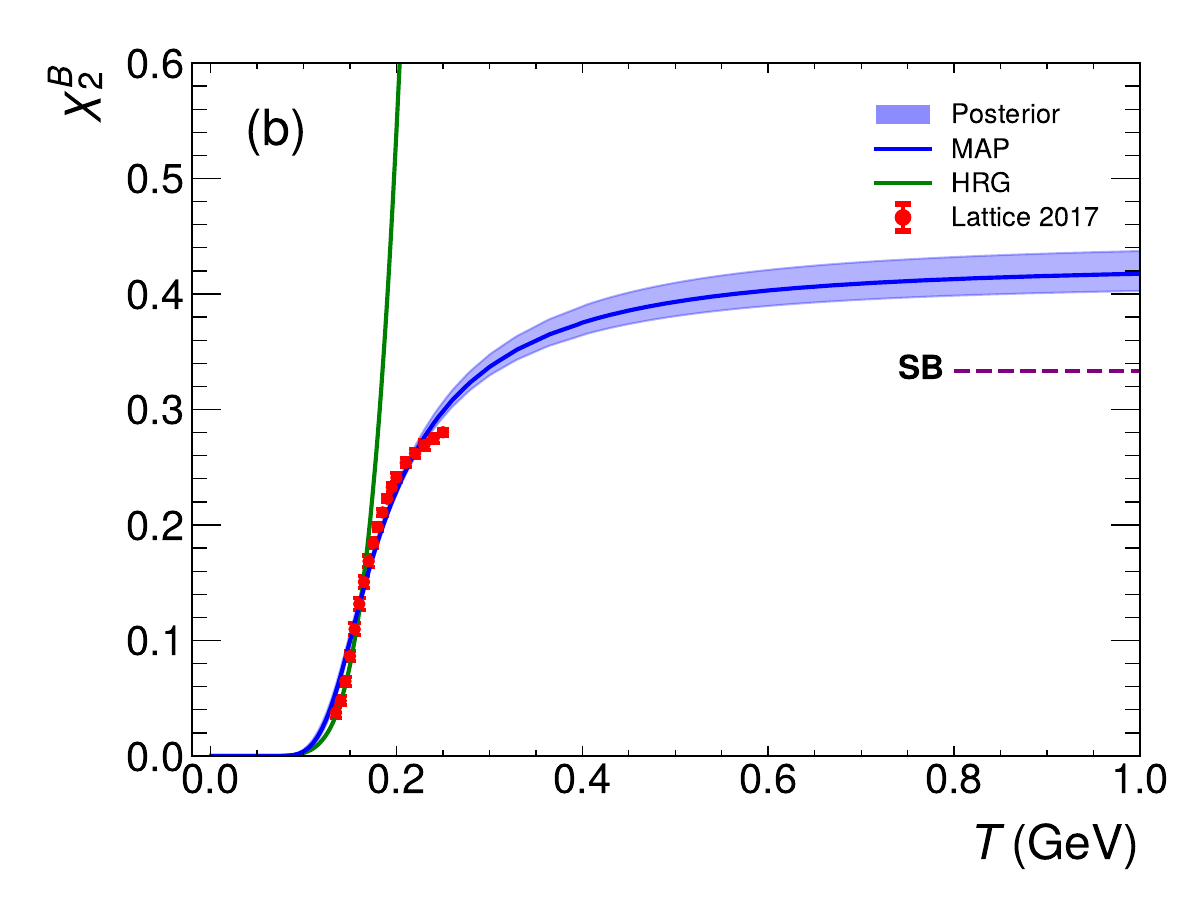}
    \end{minipage}
    \begin{minipage}{0.33\textwidth}
        \includegraphics[width=\textwidth]{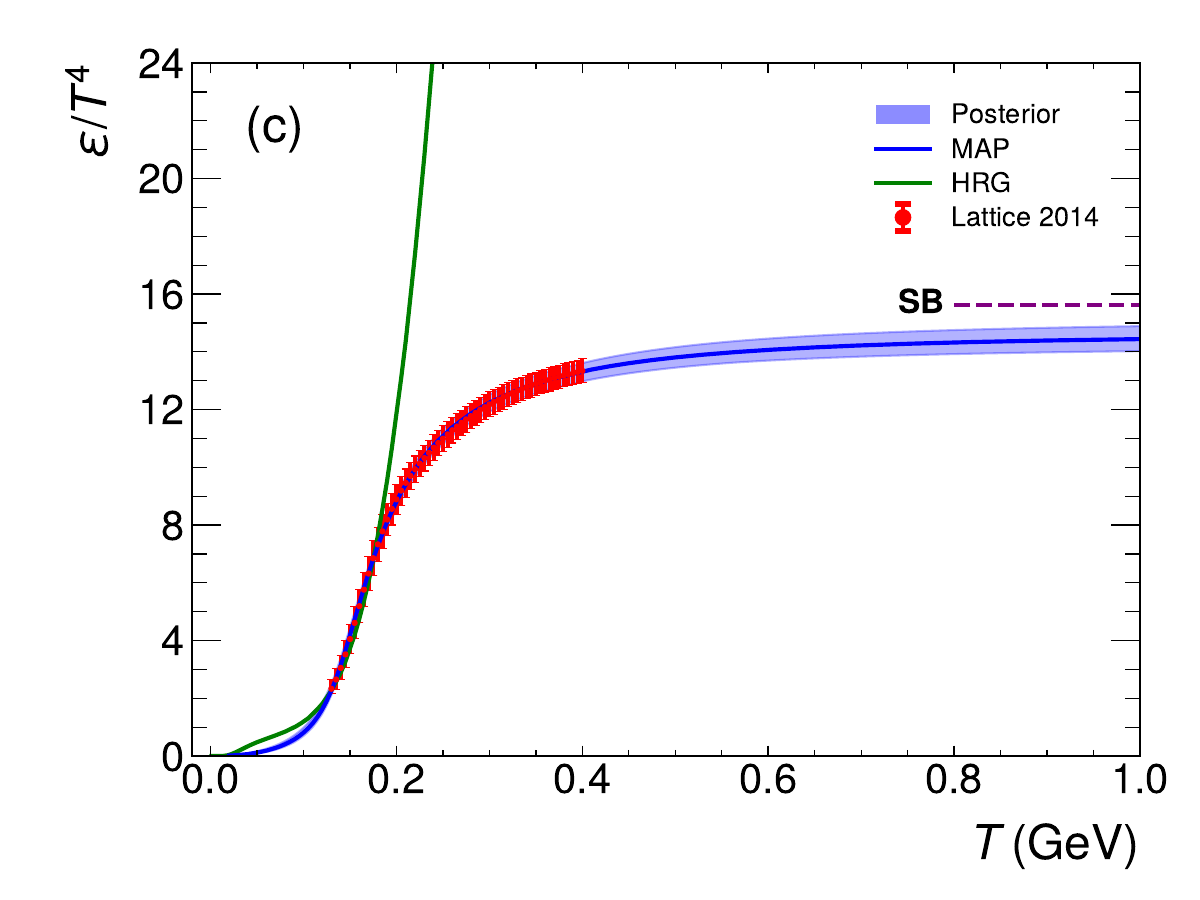}
    \end{minipage}
    \begin{minipage}{0.33\textwidth}
        \includegraphics[width=\textwidth]{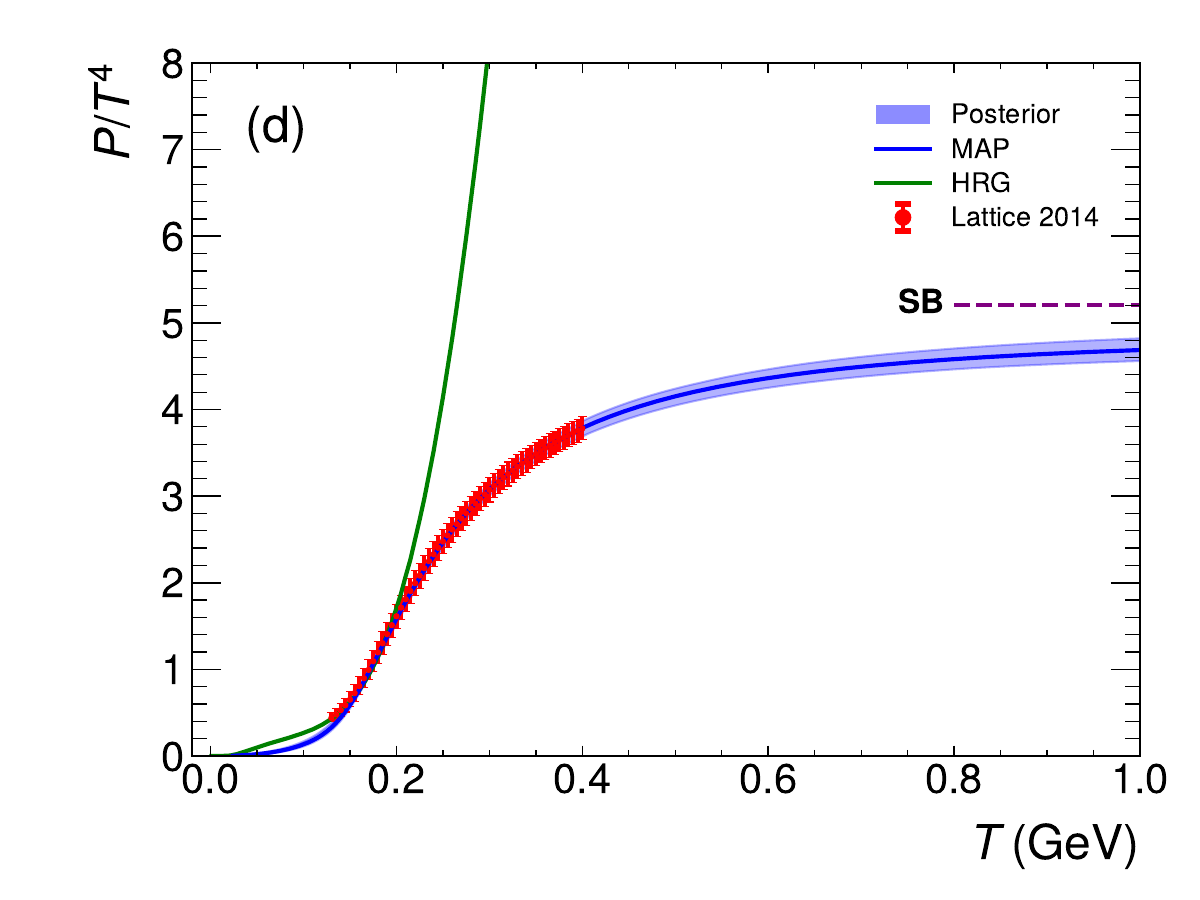}
    \end{minipage}
    \begin{minipage}{0.33\textwidth}
        \includegraphics[width=\textwidth]{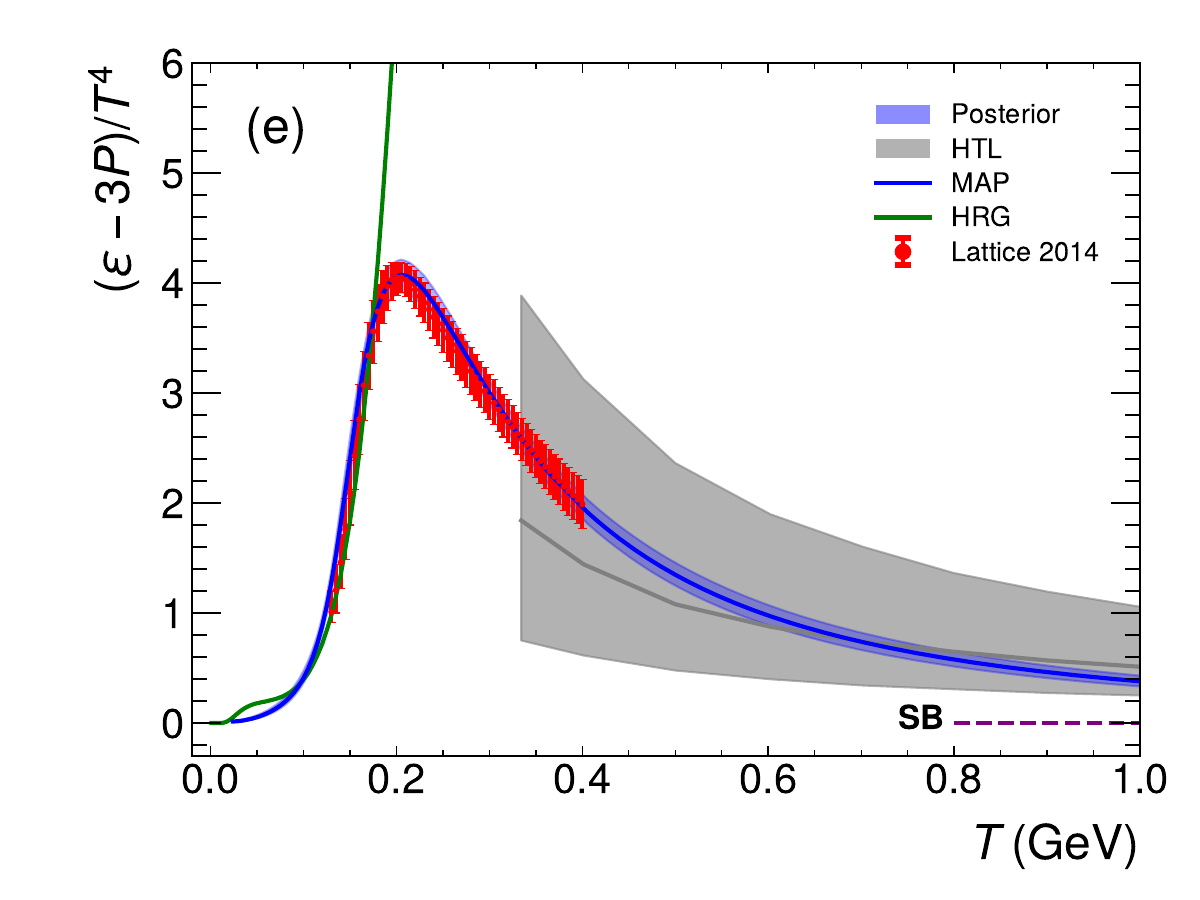}
    \end{minipage}
    \begin{minipage}{0.33\textwidth}
        \includegraphics[width=\textwidth]{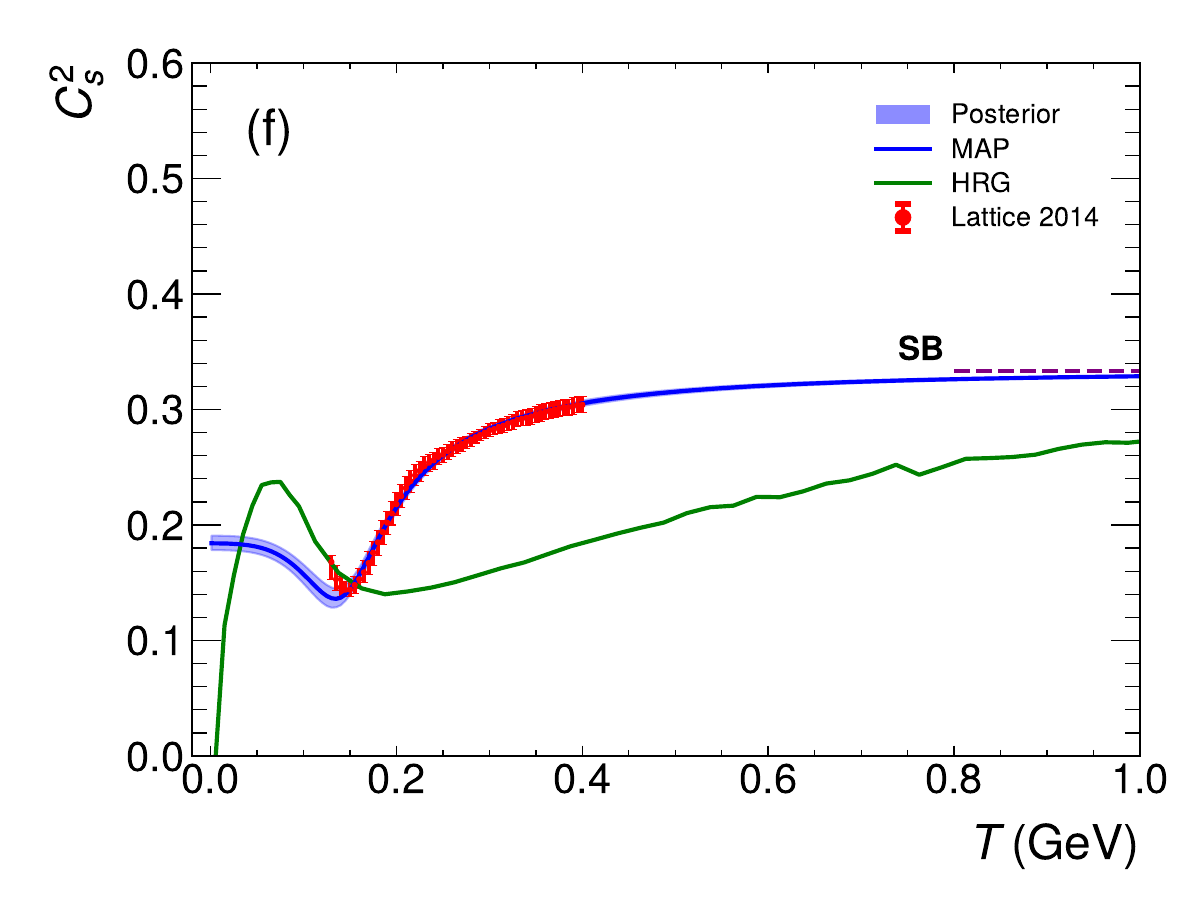}
    \end{minipage}             
    \begin{minipage}{0.33\textwidth}
        \includegraphics[width=\textwidth]{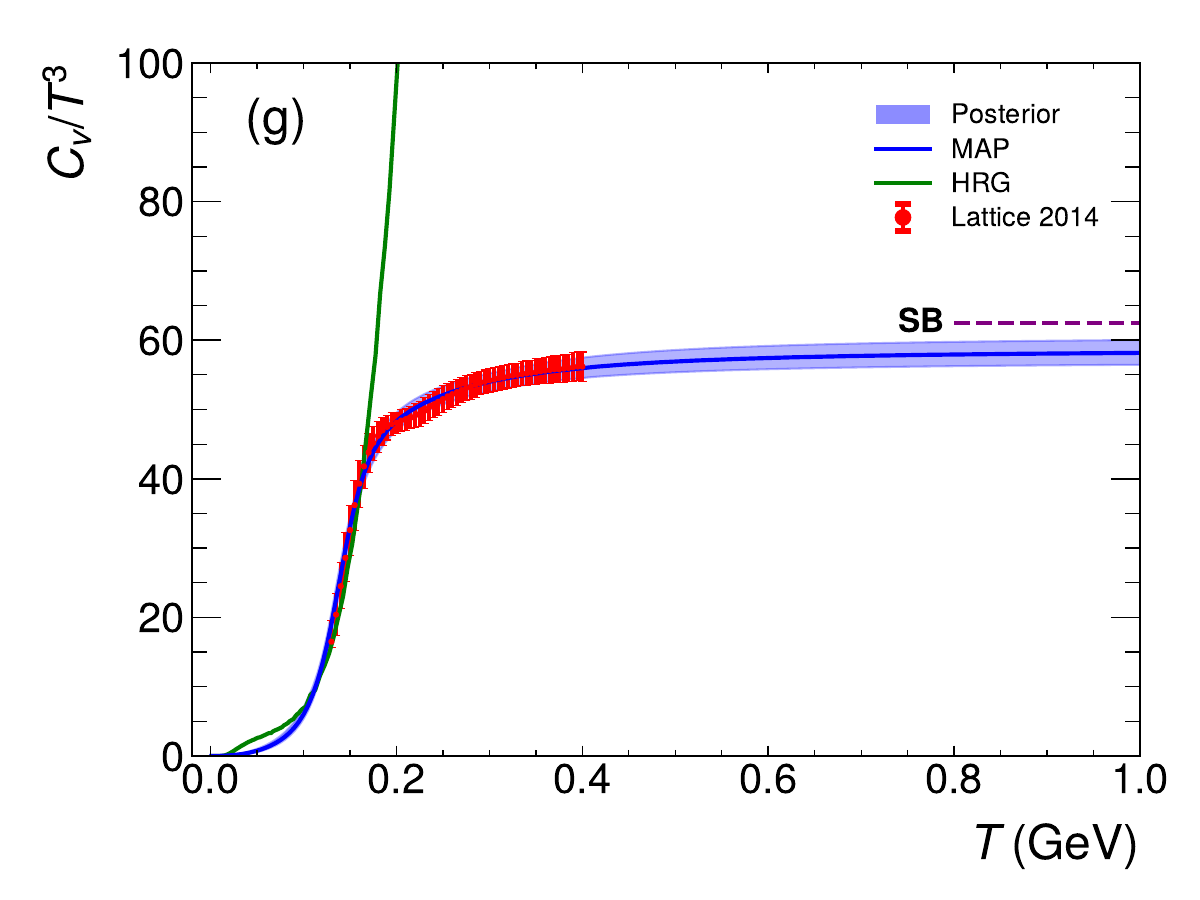}
    \end{minipage}
    \begin{minipage}{0.33\textwidth}
        \includegraphics[width=\textwidth]{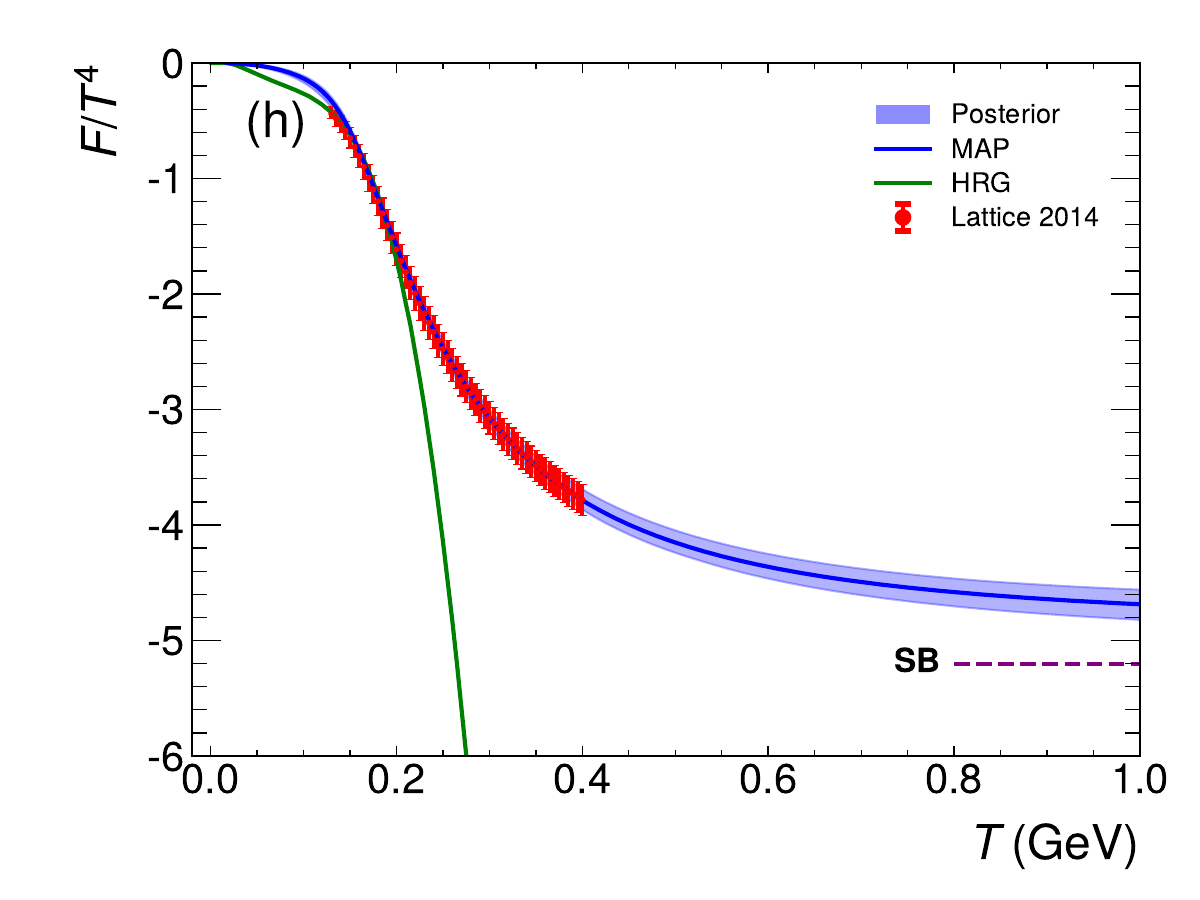}
    \end{minipage}
    \begin{minipage}{0.33\textwidth}
        \includegraphics[width=\textwidth]{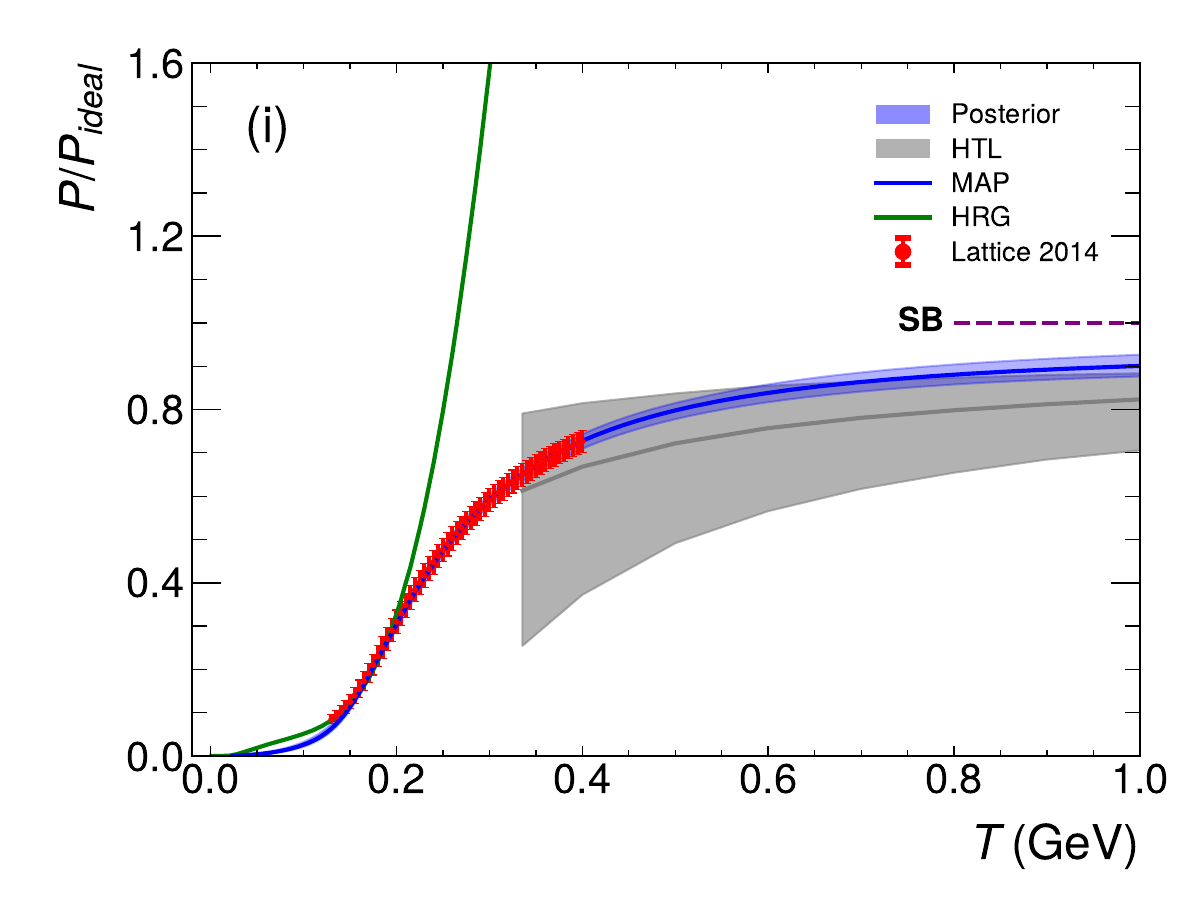}
    \end{minipage}
    \begin{minipage}{0.33\textwidth}
        \includegraphics[width=\textwidth]{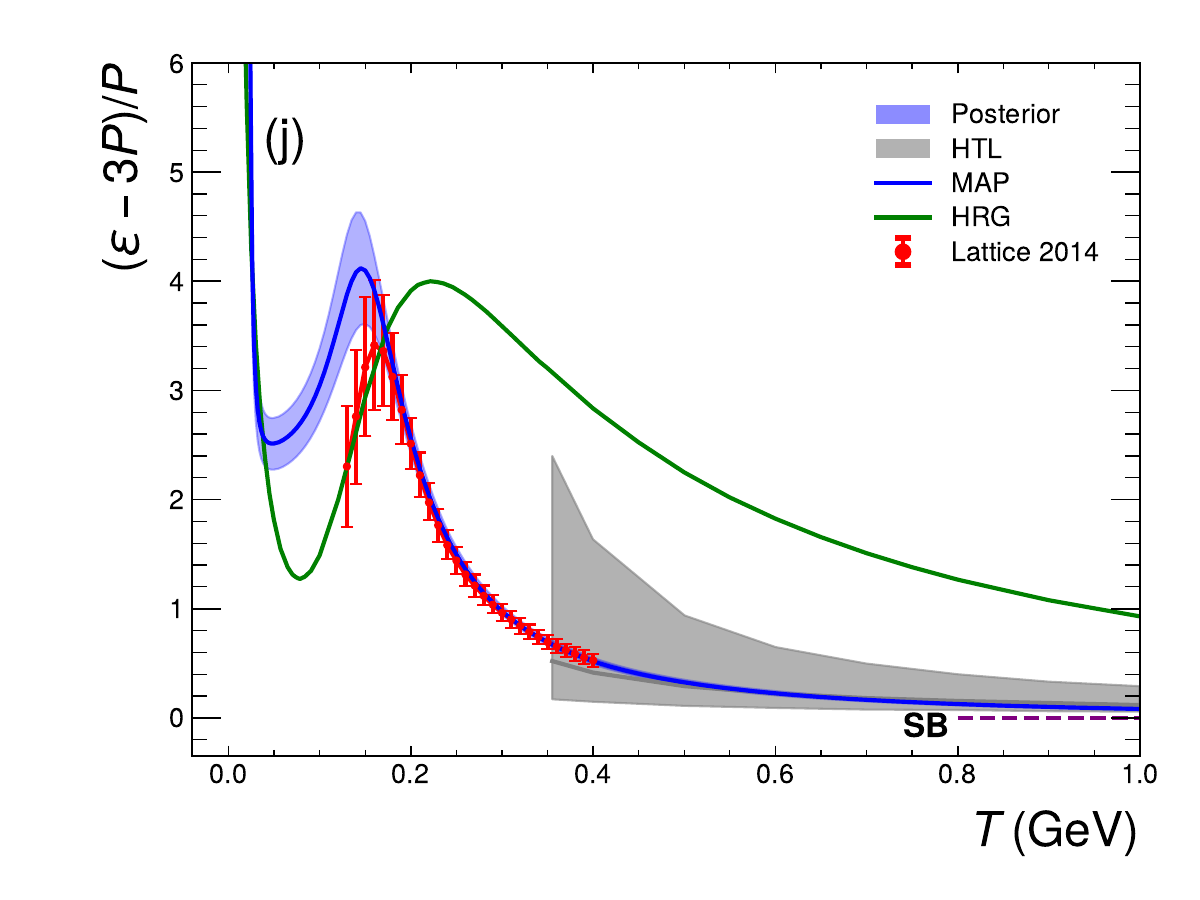}
    \end{minipage}
    \caption{Calculated the posterior results from our model (case 2, blue bands: $95\%$ confidence levels; solid lines: MAP) compared to lattice QCD, HRG and HTL predictions. Subfigures a - j correspond to the entropy, susceptibility, energy, pressure, trace anomaly, the square of the speed of sound, the specific heat, free energy, the ratio of pressure to the ideal gas limit pressure and the ratio of the trace anomaly to pressure as functions of temperature, respectively. Lattice result from Ref. \cite{HotQCD:2014kol, Bazavov:2017dus}, HRG results from Ref. \cite{Vovchenko:2019pjl}, HTL results from Ref. \cite{Haque:2014rua}.}
    \label{fig6}
\end{figure*}

\subsection{Evidence Selection and Thermodynamic Consistency} \label{sec4.1}

By applying the same Bayesian inference procedures as described above, but with two different cases for evidence, we evaluated the corresponding MAP values for the model parameters, and further derive the thermodynamics by incorporating the EMD model on these parameters constraints. 

Key thermodynamic observables using MAP values from case 1 (green curves) and case 2 (blue curves) are displayed in Fig.\ref{fig5}. 
We see that our model within case 2 yield better agreement with lattice data compared to case 1. Specifically, for entropy ($S/T^3$) and baryon susceptibility ($\chi^B_2$), both cases align with lattice data at $T>150$ MeV, but case 2 results better captures the inflection near $T\sim 140$ MeV, where $\chi^B_2$ transitions toward the HRG limit (Fig. \ref{fig5} a - b). For the square of the speed of sound squared ($C^2_s$), while case 1 underestimates the lattice result in the region $T\sim 150$ MeV, case 2 results resolves the minimum region much better (Fig. \ref{fig5} f), reproducing the lattice simulated smooth crossover reflected by several different thermodynamics. Also obsiously we see that the inclusion of $C^2_s$ inside the evidence (case 2) reduces deviations from lattice results on the trace anomaly ($(\epsilon-3P)/T^4$) especially in the transition region. These improvements significantly enhance the consistency of the final results with the lattice data, highlight the necessity of including $C^2_s$ within the evidence to refine our Bayesian calibration for the EMD model. We thus adopt case 2 for all subsequent analysis.

\subsection{Comparison to HRG and HTL} \label{sec4.2}

In this subsection, we evaluate both the low and high temperature thermodynamics from our Bayesian Holographic model, and confront them (within case 2 analysis) with HRG and HTL estimations. We used the C++ based software package Thermal-FIST \cite{Vovchenko:2019pjl} to calculate the results of the HRG model. Thermal-FIST is a tool specifically designed for studying heavy-ion collisions and hadronic equations of state. In addition, we incorporated the HTL results \cite{Haque:2014rua} for comparison, which provides perturbative estimation over thermodynamics at the high-temperature region. 

Fig. \ref{fig6} present posterior results from our model (case 2, blue bands: $95\%$ confidence levels; solid lines: MAP) compared to lattice QCD, HRG and HTL predictions. In the figures, SB refers to the Stefan Boltzmann limit.
We see that at intermediate temperatures ($150$ MeV $< T < 400$ MeV), all observables from our model align well with lattice data within uncertainties, with the speed of sound square $C^2_s$ approaching the conformal limit $1/3$ roughly at $T>300$ MeV. At low temperature regime ($100$ Mev $< T < 150$ MeV, the calibrated holographic model aligns with HRG predictions for $\epsilon/T^4, P/T^4$ and $(\epsilon-3P)/T^4$,

\noindent while the agreement with lattice data is slightly worse. At high-temperature regime ($T>400$ MeV), the model results well lie within the uncertainty band of HTL calculations for the trace anomaly $(\epsilon-3P)/T^4$ and pressure ratio $P/P^{\text{ideal}}$, validating the perturbative asymptotic QCD behavior out of our Bayesian Holographic model, though not reaching the SB limit for the chosen temperature range (except for $\chi^B_2$). It's worth noting that the evidence from lattice QCD in the Bayesian inference only cover temperature region $125$ MeV $< T < 400$ MeV. Overall, Fig. \ref{fig6} indicates that our calibrated holographic model effectively embedded the lattice EoS and generalize well to both low and high temperature with the HRG and HTL predictions automatically matched.

Notably, the specific heat ($C_V$) and free energy ($F$) exhibit single-valued monotonicity, supporting a smooth crossover transition without discontinuities. Minor deviations in $\chi^B_2$ at around $150 - 250$ MeV and its exceeding to the SB limit indicate refinements to the holographic model setup (e.g., $f(\phi)$) could be warranted for future study.

\begin{figure*}[tbp]
    \centering
    \begin{minipage}{0.476\textwidth}
        \centering
        \includegraphics[width=\textwidth]{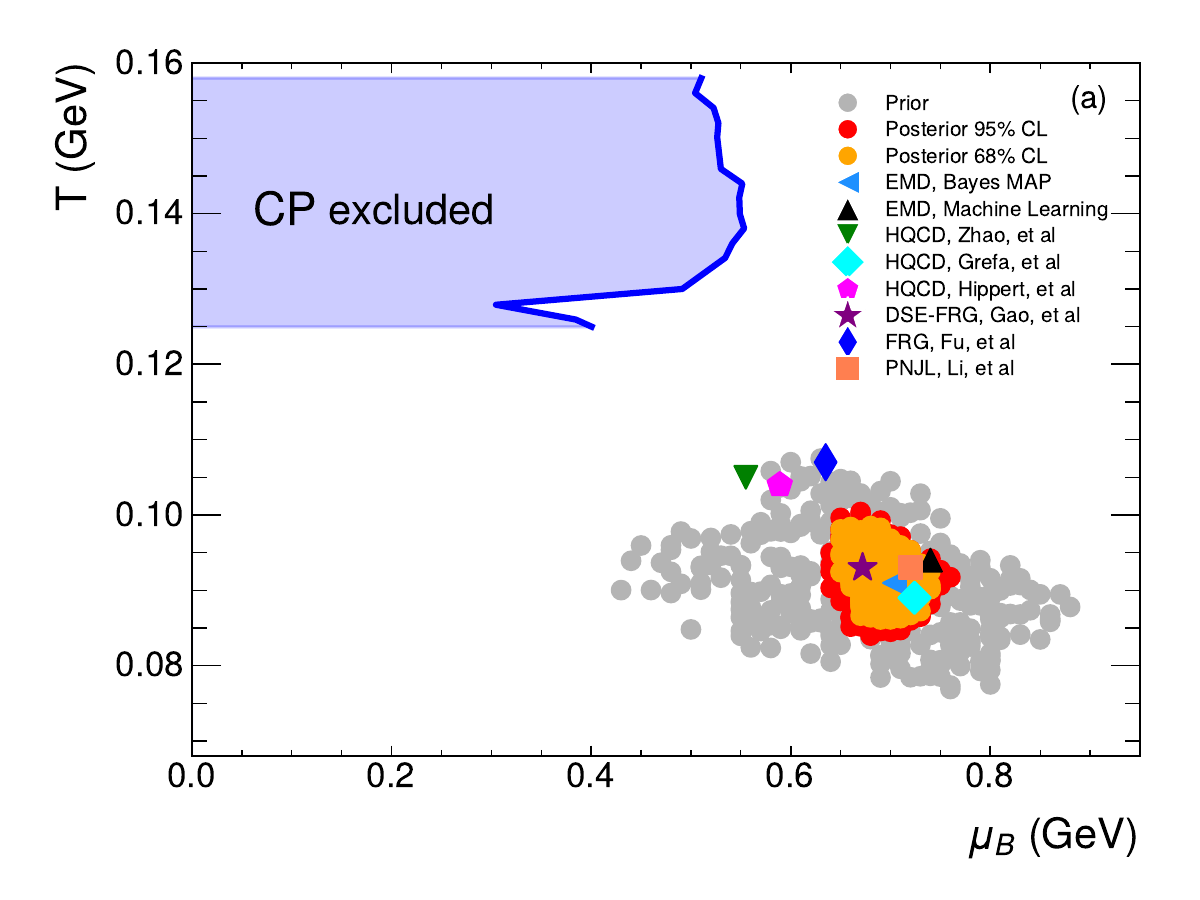}
    \end{minipage}\hfill
    \begin{minipage}{0.476\textwidth}
        \centering
        \includegraphics[width=\textwidth]{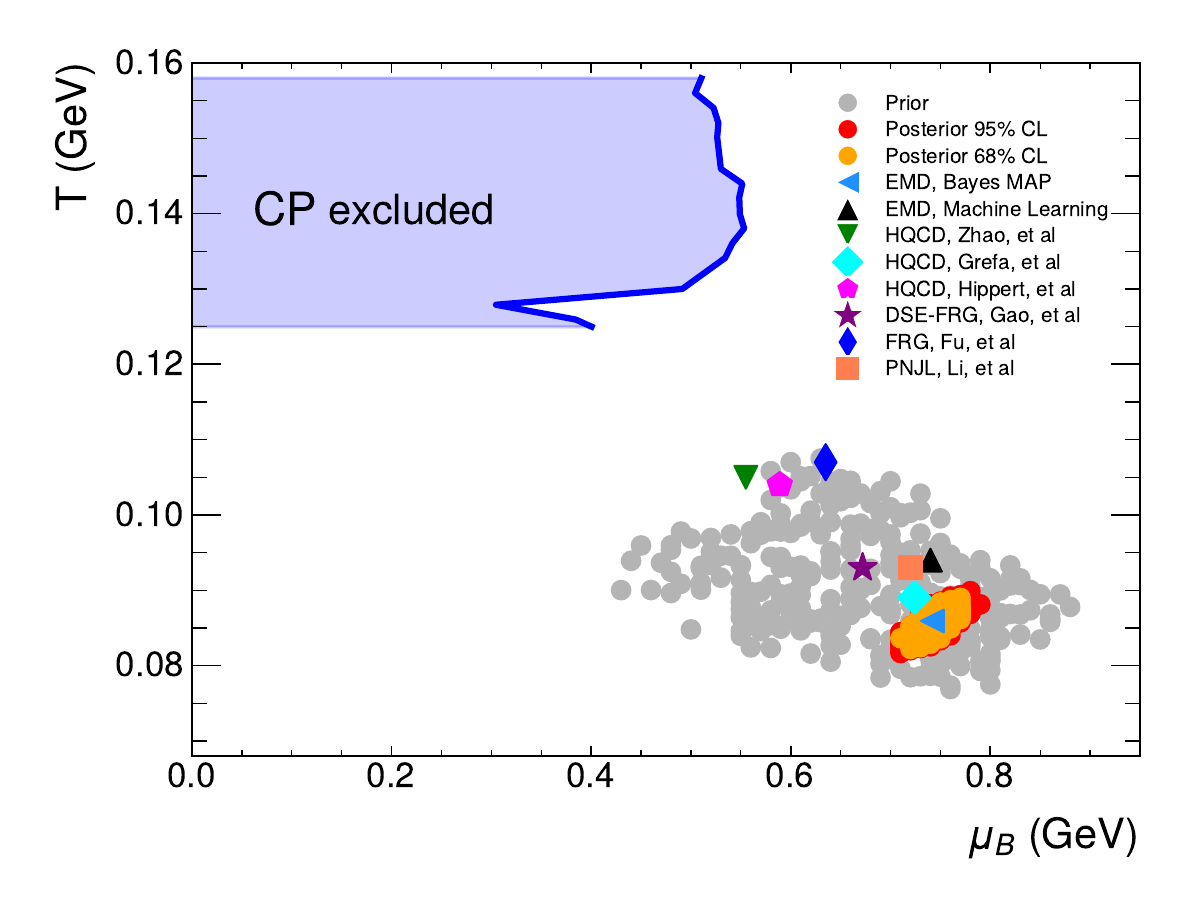}
    \end{minipage}
    \caption{CEP locations for initial prior samples (gray dots) and posterior samples (red dots at the 95\% confidence level and orange dots at the 68\% confidence level), The blue triangle indicates MAP result. Panel (a) shows the results obtained when using  $S/T^{3}$ and $\chi_{2}^{B}$ as evidence (case 1), while (b) displays those obtained with  $S/T^{3}$, $\chi_{2}^{B}$ and $C_{s}^{2}$ as evidence (case 2). The black upper triangles are results from the machine learning assisted EMD model \cite{Chen:2024mmd}. The green lower triangle \cite{Zhao:2022uxc}, cyan diamond \cite{Grefa:2021qvt}, and magenta pentagon \cite{Hippert:2023bel} denote predictions from DGR-type holographic QCD models. The purple star, the 
   blue diamond, and the coral square are predictions from DSE-FRG \cite{Gao:2020qsj}, FRG \cite{Fu:2019hdw} and realistic PNJL model \cite{Li:2018ygx}, respectively. CEP does not exist in the blue region \cite{Borsanyi:2025dyp}.}
    \label{fig7}
\end{figure*}

\section{QCD Phase diagram in the Beyesian holographic model} \label{sec5}

In this section, we summarize the posterior distribution of the CEP predicted by the model under the two aforementioned evidence cases based on Bayesian inference, as shown in Fig. \ref{fig7}. The location of the CEP ~\cite{DeWolfe:2010he,He:2013qq,Yang:2014bqa,Yang:2015aia,Cai:2022omk} is determined by finding the maximum of the gradient of entropy with respect to temperature at zero chemical potential.

Fig.\ref{fig7} displays the CEP locations in the $(T,\mu_B$ plane for both prior (gray) and posterior (red/orange for $95\%$/$68\%$ confidence level) samples, as well as for the MAP (blue triangle).
Panel (a) corresponds to case 1 where only $S/T^3$ and $\chi^B_2$ serves as evidence in the Bayesian analysis. The posterior $68\%$ and $95\%$ confidence intervals for the critical temperature and chemical potential are $\mu_{B}^{c}$ are $(T^{c},\; \mu_{B}^{c})_{68\%}$=$(0.0846\text{--}0.0985,\; 0.65\text{--}0.74)$ GeV and $(T^{c},\; \mu_{B}^{c})_{95\%}$=$(0.0839\text{--}0.1003,\; 0.64\text{--}0.76)$ GeV,  respectively, and the MAP result is ($T^{c}$=$0.0909$ GeV, $\mu_{B}^{c}$=$0.704$ GeV). 
Panel (b) presents the scenario in which $S/T^3$, $\chi^B_2$ and $C^2_s$ are included in the evidence. The 68\% and 95\% confidence intervals for $(T^{c},\mu_{B}^{c})$ become: $(T^{c},\mu_{B}^{c})_{68\%}$=$(0.0820\text{--}0.0889,\; 0.71\text{--}0.77)$ GeV and $(T^{c},\; \mu_{B}^{c})_{95\%}$=$(0.0816\text{--}0.0898,\; 0.71\text{--}0.79)$ GeV, respectively, and the MAP result is ($T^{c}$=$0.0859$ GeV, $\mu_{B}^{c}$=$0.742$ GeV).

Additionally, we compared these results with various other CEP predictions from different theoretical frameworks, such as DSE-FRG \cite{Gao:2020qsj}, FRG \cite{Fu:2019hdw}, the realistic PNJL model \cite{Li:2018ygx}, and the machine learning method based on EMD \cite{Chen:2024mmd}. 
We observed that the machine learning assissted EMD model in previous work\cite{Chen:2024mmd} falls within the $95\%$ confidence interval of the Bayesian holographi model when $S/T^3$ and $\chi^B_2$ alone are considered evidence [Panel(a)]. However, it lies outside the posterior prediction in Panel(b), likeliy because the machine learning analysis didn't incorporate $C^2_s$ explicitly as a constraint.
Compared to case 1, case 2 incorporates additional the squared speed of sound data as a prior on the basis of case 1. Technically, this leads to differences in the fitting performance of the holographic EMD model to lattice data. Case 2 achieves a better fit to the lattice data. The reason for this difference is that the EMD model we use cannot fully represent the physics of lattice QCD, meaning the accuracy of the fitting cannot be guaranteed, our calculations show that when more constraints are added (case 2: additional inclusion of $C_{s}^{2}$ data), the fitting accuracy of the EMD model to lattice data improves (as shown in Fig. \ref{fig5}), however, the predicted location of the CEP depends on the accuracy of the model's fit to the lattice data. Therefore, incorporating the squared speed of sound data has a significant impact on the predicted location of the CEP.
Finally, it should be noted that our current calculations and predictions are model dependent. 
Any modification to the holographic EMD model would consequently alter the Bayesian inference results, posterior calculations, and CEP predictions.
The uncertainty in the final results is not only related to the errors in the input lattice data but also to the errors from the Gaussian emulator during the Bayesian inference process. We have considered both factors in our calculations. At the same time, the uncertainty of our holographic EMD model may also affect the final results. The uncertainty in the holographic EMD model arises from the assumed forms of the two functions $f(z)$ (Eq. \ref{fff} of the paper) and $A(z)$ (Eq. \ref{AAA} of the paper). Different assumed forms will lead to minior changes in the final calculation results. We will attempt to address how to incorporate the model's uncertainty into the calculations in our future work.

\section{Conclusion} \label{sec6}
Building on previous work \cite{Chen:2024ckb,Chen:2024mmd}, in this paper, we systematically integrate the error estimates from lattice QCD data into a Bayesian holographic model, leveraging Bayesian inference to evaluate the QCD thermodynamics and predict the location of the CEP. In this model, we employ $S/T^{3}$, $\chi_{2}^{B}$ and $C_{s}^{2}$ lattice data as evidence to perform Bayesian inference, yielding a MAP estimate for the CEP at ($T^{c}$=$0.0859$ GeV, $\mu_{B}^{c}$=$0.742$ GeV). We also provide the corresponding uncertainties for our estimation over the positions of the CEP within both 68\% and 95\% confidence levels. The 68\% and 95\% confidence levels for the critical temperature $T^{c}$ and chemical potential $\mu_{B}^{c}$ are $(T^{c},\; \mu_{B}^{c})_{68\%}$=$(0.0820\text{--}0.0889,\; 0.71\text{--}0.77)$ GeV and $(T^{c},\; \mu_{B}^{c})_{95\%}$=$(0.0816\text{--}0.0898,\; 0.71\text{--}0.79)$ GeV. Additionally, we used the $C_{s}^{2}$ criterion to calculate the temperature range of the smooth crossover at zero chemical potential when $C_{s}^{2}$ reaches its minimum in the posterior distribution. The temperature ranges corresponding to the 95\% confidence levels is found to be $T_{95\%}$=$0.1297\text{--}0.1386$ GeV. Our model's predictions are consistent with results of established theoretical frameworks in their corresponding valid temperature range. 
From the perspective of CEP predictions, we do have some differences compared to previous predictions. In our calculations, case 2 demonstrates better fitting performance for the lattice data of thermodynamic quantities overall (as shown in Figs. \ref{fig5}-\ref{fig6}), and the model's CEP position depends on the fitting performance of the lattice data (as shown in Fig. \ref{fig7}).
In the future, we plan to extend our model to calculate additional physical properties for QCD matter, including the in-medium heavy-quark potential~\cite{Luo:2024iwf}, diffusion coefficient, and viscosity, to further validate our model's accuracy and applicability.

\section{Acknowledgement}
This work is supported in part by the National Natural Science Foundation of China (NSFC) Grant Nos: 12405154, 12235016, 12221005 and the Strategic Priority Research Program of Chinese Academy of Sciences under Grant No XDB34030000, the Fundamental Research Funds for the Central Universities, Open fund for Key Laboratories of the Ministry of Education under Grants No.QLPL2024P01, CUHK-Shenzhen university development fund under grant No.\ UDF01003041 and UDF03003041, Shenzhen Peacock fund under No.\ 2023TC0179, and the European Union -- Next Generation EU through the research grant number P2022Z4P4B ``SOPHYA - Sustainable Optimised PHYsics Algorithms: fundamental physics to build an advanced society'' under the program PRIN 2022 PNRR of the Italian Ministero dell'Universit\`a e Ricerca (MUR).

\bibliography{ref_temp}
\end{document}